\documentclass[a4paper,11pt]{article}
\pdfoutput=1 % if your are submitting a pdflatex (i.e. if you have
\usepackage{jheppub} % for details on the use of the package, please
\usepackage{amsmath,amssymb,bm,bbm}
\usepackage[utf8]{inputenc}
\renewcommand{\gg}{\mathfrak{g}}
\renewcommand{\sl}{\mathfrak{sl}}
\newcommand{\cW}{\mathcal W}

\title{Correlator correspondences for subregular  \(\mathcal{W}\)-algebras and principal \(\mathcal{W}\)-superalgebras}

\author[a]{Thomas Creutzig}
\author[b]{Yasuaki Hikida}
\author[a]{and Devon Stockal}

\affiliation[a]{Department of Mathematical and Statistical Sciences, University of Alberta, Edmonton, \\Alberta T6G 2G1, Canada}
\affiliation[b]{Center for Gravitational Physics, Yukawa Institute for Theoretical Physics, Kyoto University, \\Kyoto 606-8502, Japan}

\emailAdd{creutzig@ualberta.ca}
\emailAdd{yhikida@yukawa.kyoto-u.ac.jp}
\emailAdd{stockall@ualberta.ca}
\abstract{
We examine a strong/weak duality between a Heisenberg coset of a theory with  $\mathfrak{sl}_n$ subregular $\mathcal{W}$-algebra symmetry and a theory with a $\mathfrak{sl}_{n|1}$-structure. In a previous work, two of the current authors provided a path integral derivation of correlator correspondences for a series of generalized Fateev-Zamolodchikov-Zamolodchikov (FZZ-)duality. In this paper, we derive correlator correspondences in a similar way but for a different series of generalized duality. This work is a part of the project to realize the duality of corner vertex operator algebras proposed by Gaiotto and Rap\v{c}\'ak and partly proven by Linshaw and one of us in terms of two dimensional conformal field theory. We also examine another type of duality involving an additional pair of fermions, which is a natural generalization of the fermionic FZZ-duality. The generalization should be  important since a principal $\mathcal{W}$-superalgebra appears as its symmetry and the properties of the superalgebra are less understood than bosonic counterparts.

 }

\arxivnumber{2106.15073}
\preprint{YITP-21-68}

\begin{document}
	\maketitle
	\flushbottom

%%%%%%%%%%%%%%%%%%%%%%%%%%%%%%%%%%%%%%%%%%%%%%%%%%%%%%%%%%%%%%%%%%%%%%

\section{Introduction and summary}

In general, strong/weak duality in quantum field theory is quite important since it allows us to examine strongly coupled physics from a weakly coupled theory. In this paper, we examine a series of strong/weak dualities in two dimensional conformal field theory (CFT) as exactly tractable examples. A fundamental example may be given by the Fateev-Zamolodchikov-Zamolodchikov (FZZ-)duality \cite{FZZ}. It is an equality between Witten's two dimensional cigar model \cite{Witten:1991yr} described by a coset $SL(2)/U(1)$ with level $k$ and Sine-Liouville theory with two bosons $\phi,X$ and interaction terms $e^{\sqrt{k-2} \phi} \cos (\sqrt{k} X)$. 
In \cite{Hikida:2007tq},  a path integral derivation of the relation among correlation functions of 
$\mathfrak{sl}_2$ Wess-Zumino-Novikov-Witten (WZNW) model and Liouville field theory   \cite{Ribault:2005wp,Ribault:2005ms} was given. 
It led to a proof of the FZZ-duality   in \cite{Hikida:2008pe} by applying the technique developed in \cite{Hikida:2007tq} along with the self-duality of Liouville field theory.
There is also a fermionic version of the FZZ-duality proven in \cite{Hori:2001ax} as a mirror symmetry and  in \cite{Creutzig:2010bt} by applying the method of  \cite{Hikida:2007tq}.
These dualities have been applied to examine holography, singular Calabi-Yau manifolds and so on, see. e.g.,  \cite{Kazakov:2000pm,Ooguri:1995wj,Giveon:1999zm}.

It is desired to have more dualities that resemble  the FZZ-duality.
In a series of papers \cite{Hikida:2007sz,Creutzig:2011qm,Creutzig:2015hla}, the reduction method of \cite{Hikida:2007tq} has been generalized for many examples, and particularly, in a recent paper \cite{Creutzig:2020ffn}, the range of applicability was significantly enlarged.
Applying the result and the self-duality of Toda field theory, a series of generalized FZZ-dualities was derived by two of the current authors in \cite{Creutzig:2020cmn}. This duality is between a higher rank coset $SL(n+1)/(SL(n) \times U(1))$ and a theory with a $\mathfrak{sl}_{n+1|n}$-structure and it has a nature of strong/weak duality as in the case of FZZ-duality. The symmetry algebras underlying the generalized FZZ-dualities are (truncations) of $\mathcal{W}_\infty$-algebra, and the importance of $\mathcal{W}$-algebras in general has  increased significantly in this decade. For instance, the subsectors of four dimensional supersymmetric gauge theories are known to be organized by $\mathcal{W}$-algebras \cite{Alday:2009aq,Wyllard:2009hg}. Moreover, the coset $SL(n+1)/(SL(n) \times U(1))$ is essentially the one appearing in the higher spin holography proposed in \cite{Gaberdiel:2010pz}  or its supersymmetric version in \cite{Creutzig:2011fe}.%
\footnote{It would be useful to introduce extended supersymmety in order to examine possible relations between superstring theory and higher spin gravity as in \cite{Gaberdiel:2013vva,Gaberdiel:2014cha} with $\mathcal{N}=4$ supersymmetry  or as in \cite{Creutzig:2011fe,Creutzig:2013tja,Creutzig:2014ula} with $\mathcal{N}=3$ supersymmetry.}

More precisely speaking, the symmetry algebras for the dualities can be identified with the corner vertex operator algebras (VOAs) constructed by \cite{Gaiotto:2017euk}. The VOAs can be realized via brane junctions and denoted by $Y_{n_1 , n_2 , n_3}[\psi]$, where $n_1,n_2,n_3$ are non-negative integers and $\psi$ is related to the coupling constant of the algebra. The duality of corner VOAs can be deduced from the string duality of the brane junctions,
and in particular, a strong/weak duality is obtained by exchanging $n_1$ and $n_2$ as $Y_{n_1,n_2,n_3}[\psi] \simeq Y_{n_2,n_1,n_3}[\psi ^{-1}] $. See the next subsection for more mathematical explanations. 
Since the symmetry of the coset $SL(n+1)/(SL(n) \times U(1))$ is realized by $Y_{0,n,n+1}$-algebra,
the higher rank FZZ-duality in  \cite{Creutzig:2020cmn} can be regarded as a CFT realization of the duality $Y_{0,n,n+1}[\psi] \simeq Y_{n,0,n+1}[\psi ^{-1}]$. Indeed, it was found that the interaction terms in the dual theory correspond to the screening operators in a (conjectured) free field realization of $Y_{n,0,n+1}$-algebra in \cite{BFM, Litvinov:2016mgi,Prochazka:2018tlo}.

In this paper, we examine another series of strong/weak dualities between a Heisenberg coset of a theory with the symmetry of $\mathfrak{sl}_n$ subregular $\mathcal{W}$-algebra and a theory with a $\mathfrak{sl}_{n|1}$-structure. The symmetry of the coset is given by the $Y_{0,1,n}$-algebra, while the dual theory corresponds to the free field realization of $Y_{1,0,n}$-algebra obtained in appendix \ref{sec:Yalgebra}. Thus the duality corresponds to that of VOAs $Y_{0,1,n}[\psi] \simeq Y_{1,0,n}[\psi ^{-1}] $, and the current work can be regarded as a part of the project to realize the duality of corner VOAs in terms of conformal field theory. We derive the duality by applying the analysis of \cite{Hikida:2008pe,Creutzig:2020cmn} with the reduction method of \cite{Hikida:2007tq,Creutzig:2020ffn} particularly.
We also examine another series of duality by adding a pair of fermions, which can be regarded as a generalization of fermionic FZZ-duality mentioned above.
In this case, the symmetry algebra is given by the regular $\mathcal{W}$-superalgebra of $\mathfrak{sl}_{n|1}$.
This version of duality is useful to examine the properties of $\mathcal{W}$-superalgebra from those of bosonic counterparts \cite{Creutzig:2020vbt,Creutzig:2021bmz}.

In this paper, we only examine correlation functions on a Riemann sphere.
However, it would be important to generalize the analysis in the case with Riemann surfaces of higher genus and/or with boundaries. For example, the original FZZ-duality was extended to the case of higher genus Riemann surface in \cite{Hikida:2008pe} and the disk correlators were analyzed in \cite{Creutzig:2010bt}.

\subsection{$\mathcal W$-algebras and triality}

The general construction of $\cW$-superalgebras has been first rigorously introduced by Kac and Wakimoto \cite{Kac:2003jh}.
Let $\gg$ be a simple Lie superalgebra and $k$ a complex number. Then the symmetry algebra of the WZNW theory of $\gg$ at level $k$ is the simple affine vertex superalgebra of $\gg$ at level $k$. Here we use the physics convention of the level which differs from the mathematics convention by a minus sign. 
Let  $f$ be an even nilpotent element in $\gg$. Then one can add ghosts to the affine vertex superalgebra and a BRST-charge $Q$. Both are determined by the choice of nilpotent element. The corresponding BRST-cohomology is then the $\cW$-algebra of $\gg$ at level $k$ associated to $f$. The corresponding conformal field theory is of Toda-type. For example if $\gg = \sl_2$ and $f$ principal (i.e. non-zero), then the $\cW$-algebra is the Virasoro algebra of central charge $13-6(b^2 + b^{-2})$ with $b^2 = k-2$. The conformal field theory is Liouville theory. 

The best-known $\cW$-algebras are those where $\gg$ is a Lie algebra and $f$ is principal nilpotent. These are the \emph{usual} $\cW$-algebras. 
For example for $\gg = \sl_n$ these are the $\cW_n$-algebras of type $2, 3, \dots, n$. The maybe second most prominent $\cW$-algebras are the subregular $\cW$-algebras of $\sl_n$. These are also called Feigin-Semikhatov algebras \cite{Feigin:2004wb}  and the case of $n=2$ is just the affine vertex algebra of $\sl_2$ and the case of $n=3$ is the Bershadsky-Polyakov algebra \cite{Bershadsky:1990bg,Polyakov:1989dm}. These algebras are of type $1, 2, \dots, n-1$ together with two more fields of conformal weight $\frac{n}{2}$. 

Triality refers to three different realizations of the same $\cW$-algebra or its coset. For example, let $k$ and $\ell$ be related by $(k-n)(\ell-n) =1$, then the $\cW_n$-algebras at level $k$ and $\ell$ are isomorphic. This is the self-duality of principal $\cW$-algebras, the Feigin-Frenkel duality \cite{FF}. 
The third member of this triality is the GKO-coset construction of principal $\cW$-algebras \cite{Goddard:1986ee}, that has been finally proven in \cite{Arakawa:2018iyk}. 

Such trialities have been conjectured by Gaiotto and Rap\v{c}\'ak  for large classes of cosets of $\cW$-superalgebras \cite{Gaiotto:2017euk} as trialities of corner VOAs mentioned above. These conjectures have been proven by Linshaw and one of us in all the cases where at least one of the involved $\cW$-superalgebras is in fact a $\cW$-algebra \cite{Creutzig:2020zaj, Creutzig:2021dda}.
In the case of the Feigin-Semikhatov algebras, that is the subregular $\cW$-algebras of $\sl_n$, the statement is as follows. This $\cW$-algebra has a free boson as subalgebra and one can consider the coset, see \cite{W23, W24} for studies of cosets of rational theories. Let $k$ and $\ell$ be related by $(k-n)(\ell-n) =1$, then this coset of the subregular $\cW$-algebras of $\sl_n$ at level $k$ is isomorphic to the free boson coset of the principal $\cW$-superalgebra of $\sl_{n|1}$. The third member in this triality is a coset of the affine vertex superalgebra of $\sl_{n|1}$ by its affine subalgebra. 

The duality between the subregular $\cW$-algebras of $\sl_n$ and the principal $\cW$-superalgebra of $\sl_{n|1}$ has been derived in \cite{Creutzig:2020vbt} using free field realizations, i.e. screening realizations that are due to Genra \cite{Genra1}. Free field realizations are a good starting point and we can now rederive this duality using our path integral formalism and thus get a full duality of conformal field theories in the sense that we also get a precise matching of correlation functions. 

\subsection{Organization}

We derive the dualities by following the previous analysis of \cite{Hikida:2008pe,Creutzig:2010bt,Creutzig:2020cmn} by applying the reduction methods developed in \cite{Hikida:2007tq,Creutzig:2020ffn}. 
As explained above, the match of the symmetry algebras was already known for the dualities.
Therefore, we need to derive correspondences for the correlation functions of primary operators.
For this, we start from the first order formulations of the coset theories. In the next section, we construct a theory with  the symmetry of $\mathfrak{sl}_n$ subregular $\mathcal{W}$-algebra corresponding to a free field realization of the algebra.%
\footnote{Precisely speaking, we construct the action with the symmetry after twisting its energy momentum tensor. After the twisting, $(\beta , \gamma)$ fields appearing in the action have conformal dimensions $(1,0)$ and the techniques of  \cite{Hikida:2007tq,Creutzig:2020ffn} become directly applicable.  See appendix \ref{sec:twisting} for details.}
In section \ref{sec:Bduality}, we examine the duality between a Heisenberg coset of the theory with the symmetry of $\mathfrak{sl}_n$ subregular $\mathcal{W}$-algebra and a theory with a $\mathfrak{sl}_{n|1}$-structure.
In subsection \ref{BosonicGauging}, we define the coset theory by gauging a $U(1)$-symmetry.
In subsection \ref{sec:Bshifting}, we perform field redefinitions as in  \cite{Hikida:2007tq,Creutzig:2020ffn} and rewrite $N$-point functions of the coset theory in terms of higher point functions of $\mathfrak{sl}_n$ Toda field theory with an additional free bosonic field. 
In subsection \ref{sec:Bdual}, we interpret the additionally inserted fields as a part of interaction terms in order to interpret the obtained theory  as  the dual one.
We also perform the self duality of the Toda field theory, which is a key point to obtain a strong/weak duality.
In subsection \ref{sec:Bchanging}, we examine the $\mathfrak{sl}_{n|1}$-structure of the theory obtained by the procedure.
In section \ref{sec:Fduality}, we repeat essentially the same analysis to the case with additional pair of free fermions. In particular, we observe that the interaction terms of the theory dual to the fermionic coset correspond to the screening operators of principal $\mathcal{W}$-superalgebra of $\mathfrak{sl}_{n|1}$ in subsection \ref{sec:Fchanging}.
In appendix \ref{sec:twisting}, we explain the sector used in the main context with a twisted energy momentum tensor.
In appendix \ref{sec:Yalgebra}, we examine free field realizations of $Y_{1,0,n}$-algebra.

\section{Subregular \(\mathcal{W}\)-algebra of  \(\mathfrak{sl}_n\) }

\label{sln walg}

In this paper, we examine dualities involving cosets, where a part is given by a theory with the symmetry of \(\mathfrak{sl}_n\) subregular \(\mathcal{W}\)-algebra. Since the algebra is called as $\mathcal{W}^{(2)}_n$-algebra in \cite{Feigin:2004wb}, we may name the theory as $\mathcal{W}^{(2)}_n$-theory.
As explained in \cite{Creutzig:2020ffn}, we should begin with a first order formulation of the $\mathcal{W}^{(2)}_n$-theory corresponding to a free field realization of the algebra, which is introduced in this section.

The $\mathcal{W}^{(2)}_n$-algebra can be expressed by free fields $\phi_i$ $(i=1,2,\ldots, n-1)$ and a ghost system $(\beta , \gamma)$.
The operator product expansions (OPEs) are assumed to be 
\begin{align}
\phi_i (z) \phi_j (w) \sim - A_{ij} \ln |z - w|^2 \, , \quad \beta(z) \gamma (w) \sim \frac{1}{z -w} \, . \label{fbOPE}
\end{align}
Let us define the constant \(b^{-1}=\sqrt{k-n}\), and $A$ to the Cartan matrix of \(\mathfrak{sl}_{n-1}\).  
It is convenient to choose screening operators given by
\begin{equation}
S_1=\alpha \beta\overline{\beta}e^{b\phi_1}\qquad \text{and} \qquad
S_l=\alpha  e^{b\phi_l} \ \ \text{for}\  l\neq 1  \label{screening}
\end{equation}
with a constant $\alpha$.
See, e.g.,  \cite{Feigin:2004wb,Creutzig:2020vbt}. We are then prepared to present a first order formulation of the action associated to the subregular  \(\mathcal{W}\)-algebra of $\sl_n$,
\begin{equation}\label{eq: action}
\begin{split}
S^\text{sub}_k&=\frac{1}{\pi}\int d^2w \left[  \frac{1}{2}\sum_{\substack{l,j=1}}^{n-1}\partial \phi_l\overline{\partial}\phi_jA_{lj}^{-1}-\beta \overline{\partial}\gamma-\overline{\beta}\partial\overline{\gamma}+\frac{1}{4}\sqrt{g}\mathcal{R}
\left( b\phi^1 + (b + b^{-1})\sum_{l=2}^{n-1} \phi^l \right)\right]  \\ 
& \quad + \sum_{l=1}^{n-1} \frac{1}{\pi}\int d^2w S_l \, .
\end{split}
\end{equation}
Here we consider the metric 
\begin{align}
ds^2 = |\rho (z)|^2 dz d \bar z \label{metric}
\end{align}
 but  typically  set $\rho(z)=1$.
 In this case, the curvature is given by
 \begin{align}
  \sqrt{g}\mathcal{R} = - 4 \partial \overline{\partial}\ln|\rho(w)| \, . \label{curvature}
 \end{align}
We notice that \(\partial \phi_i (z)\phi_j(w)= - A_{ij}/( z-w )  \), so the fields \(\phi_1,...,\phi_{n-1}\) transform in a representation of the \(\mathfrak{sl}_{n-1}\) Cartan subalgebra.  
With this in mind, we introduce \(\phi^l\), which is  the dual field to \(\phi_l\), by
\begin{align} 
	\phi^l:= \sum_{j=1}^{n-1} A_{lj}^{-1} \phi_j \, , \quad \phi^j(z)\phi_i(w)= - \delta_{ji} \ln|z-w|^2 \, .
\end{align}
In particular, \(\phi^1\) is given by
\begin{align} \label{eq: dual}
	\phi^1:=\sum_{l=1}^{n-1}A_{1l}^{-1}\phi_l=\frac{1}{n}((n-1)\phi_1+(n-2)\phi_2+...+\phi_{n-1}) \, .
\end{align}
We define vertex operators in the \(\gamma\), \(\phi_1\) and \(\phi_l\) directions for \(\l\neq 1\) as follows: 
\begin{align}
    V^\gamma_{\mu}(z)=e^{\mu\gamma(z)-\overline{\mu}\overline{\gamma}(\overline{z})} \, , \quad 
V^1_\lambda=|\mu|^{2 \lambda}e^{b\lambda\phi_1(z,\overline{z})} \, , \quad V^l_{\lambda}=e^{b\lambda\phi_l(z,\overline{z})} \, .
\label{vertexop1}\end{align}
We will use the following notation for vertex operators, where the product over the index $l$ is implied: 
\begin{equation}
    V^\gamma_{\mu}(z)V_{\lambda_l}(z)=V^\gamma_{\mu}(z)\prod_{l=1}^{n-1}V^l_{\lambda_l}(z) \, . 
\end{equation}
We recall the following OPEs for vertex operators with \(:\beta\gamma:\)
\begin{align}
    :\beta\gamma:(z)V^\gamma_{\mu,\lambda}(w)=\mu\gamma V^\gamma_{\mu,\lambda}(w)\frac{1}{z-w} \, , \quad
 :\beta\gamma:(z)V^l_\lambda(w)=0 \, .
\end{align}
Here we have introduced the notation for the normal ordering \(:A(z) B(w):\), where the divergences coming from the limit $z \to w$ are removed.
We then calculate OPEs with vertex operators and screening operators in \eqref{XVphi} and \eqref{Xscreen}, respectively, as
\begin{equation}
    \partial\phi^1(z)V^l_{\lambda}(w)=- b\lambda V^l_{\lambda}(w)\frac{\delta_{1l}}{z-w}\label{XVphi} 
\end{equation}
and
\begin{align}
\partial \phi_l (z)S_j (w)= - b\frac{A_{lj}S_j}{z-w} \, , \quad \partial \phi^1(z) S_j(w)= - b\frac{S_j}{z-w}\delta_{1j} \, , \quad :\beta\gamma: (z) S_j (w)= - \frac{S_j}{z-w}\delta_{1j} \, .\label{Xscreen}
\end{align}
In order to later construct the desired coset theory, we define the diagonal Heisenberg field $J$, and calculate its operator product expansion with screening charges as below.  We note that $J $ commutes with all screening charges as desired;
\begin{align}
    J=:\beta\gamma:-\frac{1}{b}\partial \phi^1 \, , \quad 
J(z)S_j(w)=\frac{S_j(w)}{z-w}\left(- \delta_{1j}+\delta_{1j}\right)=0 \, . \label{Jscreening} 
\end{align}
Defining our coset with respect to this particular diagonal Heisenberg field $J$ allows us to focus our attention on \(\beta\) and \(\gamma\), as well as the one field \(\phi_1\) which is not treated symmetrically by the first order formulation of the WZNW action, while the remaining fields \(\phi_l\) are mostly `along for the ride'.  The field $J$ has norm given by 

\begin{equation}
    J(z)J(w)=:\beta(z)\gamma(z):\beta(w)\gamma(w):+\frac{1}{b^2}\partial \phi^1(z)\partial \phi^1(w)= - \left(1+\frac{1}{b^2}\frac{n-1}{n}\right)\frac{1}{(z-w)^2} \, .
\end{equation}
We define the constant 
\begin{equation}
    \eta^2=\frac{1}{b^2}\frac{n-1}{n} \, . 
\end{equation}
We will also consider one last operator: a spectral flow operator \(v^s(\xi)\), with a few effects. Firstly, in the \(\phi_l\) directions, \(v^s(\xi)\) inserts the operator \(\exp ( \frac{1}{b}s\phi^1(\xi) ) \).  In the \(\beta\) and \(\gamma\) directions, \(v^s\) will determine that path integration is performed only over \(\beta\) and \(\overline{\beta}\) with zero of order \(s\) at \(w=\xi\), and will act on the modes of \(\beta\) and \(\gamma\) by sending \(\beta_n\to \beta_{n+1}\) and \(\gamma_{n}\to \gamma_{n-1}\), see \cite{Hikida:2008pe} for more details.

For the vertex operator basis we will use, it is convenient to have $J$ in a differential form.  We calculate the OPE of $J$ and our vertex operator as
\begin{equation}
\begin{split}
J(z)V^\gamma_{\mu}(w)V_{\lambda_l}(w)&=\left[:\beta\gamma:(z)-\frac{1}{b}\partial \phi^1(z)\right]V^\gamma_{\mu}(w)\prod_{l=1}^{n-1}V^l_{\lambda_l}(w)\\
&=(\mu\gamma+\lambda_1)\frac{V^\gamma_{\mu}(w)V_{\lambda_l}(w)}{z-w}= \mu\partial_\mu \frac{V^\gamma_{\mu}(w)V_{\lambda_l}(w)}{z-w} \, . 
\end{split}
\end{equation}
Notice that \(J\) acts the same as \(\mu\partial_\mu\).  We will use the following form for vertex operators
\begin{align}
    \Phi^{\lambda_l}_{m,\overline{m}}(z)=N^{\lambda_1}_{m,\overline{m}}\int\frac{d^2\mu}{|\mu|^2}\mu^m\overline{\mu}^{\overline{m}}V^\gamma_{\mu}(z)V_{\lambda_l}(z) \, , \quad N^{\lambda_1}_{m,\overline{m}}=\frac{\Gamma(-\lambda_1+1-m)}{\Gamma(\lambda_1+\overline{m})} \, .
\end{align}
Using the differential form found previously, we calculate the OPE of this operator with the diagonal Heisenberg field $J $ as
\begin{equation}\nonumber
\begin{split}
    J(z) \Phi^{\lambda_l}_{m,\overline{m}}(w) &=N^{\lambda_1}_{m,\overline{m}}\int\frac{d^2\mu}{|\mu|^2}\mu^m\overline{\mu}^{\overline{m}}J(w)V^\gamma_{\mu}(w)V_{\lambda_l}(w)\\
    &=N^{\lambda_1}_{m,\overline{m}}\int\frac{d^2\mu}{|\mu|^2}\mu^m\overline{\mu}^{\overline{m}}\frac{1}{z-w}(\mu\partial_\mu V^\gamma_{\mu,\lambda_1}(w) )V_{\lambda_l}(w) \, . 
\end{split}
\end{equation}
Notice that this may be identified with the below since they differ by a total derivative:  
\begin{equation}\nonumber
\begin{split}
    J(z) \Phi^{\lambda_l}_{m,\overline{m}}(w) &=  N^{\lambda_1}_{m,\overline{m}}\int\frac{d^2\mu}{|\mu|^2} ( - \mu\partial_\mu \mu^m )\overline{\mu}^{\overline{m}} V^\gamma_{\mu}(w)V_{\lambda_l}(w)\frac{1}{z-w}\\
    &=-mN^{\lambda_1}_{m,\overline{m}}\int\frac{d^2\mu}{|\mu|^2}\mu^m\overline{\mu}^{\overline{m}}V^\gamma_{\mu}(w)V_{\lambda_l}(w)\frac{1}{z-w}=-m\Phi^{\lambda_l}_{m,\overline{m}}(w)\frac{1}{z-w} \, . 
\end{split}
\end{equation}
Then $J(z)$ acts on \(\Phi^{\lambda_1}_{m,\overline{m}}\) as multiplication by \(- m / ( z-w ) \).

\section{Bosonic duality}
\label{sec:Bduality}

In this section, we rewrite $N$-point function of a Heisenberg coset of the $\mathcal{W}^{(2)}_n$-theory into that of a theory with a $\mathfrak{sl}_{n|1}$-structure. In the next subsection, we define the Heisenberg coset as a product of the $\mathcal{W}^{(2)}_n$-theory and a free boson (and BRST ghosts) as in the case of $SL(2)/U(1)$ coset, see, e.g., \cite{Hikida:2008pe}.
In subsection \ref{sec:Bshifting}, we apply the methods developed in \cite{Hikida:2007tq,Creutzig:2020ffn} and reduce the $\mathcal{W}^{(2)}_n$-theory into $\mathfrak{sl}_n$ Toda field theory. We also apply a similar procedure to the extra free boson as in \cite{Hikida:2008pe}.
After the reduction procedure, we typically obtain extra insertion of fields, and in subsection \ref{sec:Bdual} we interpret them as a part of interaction terms.
During the process, we apply the self-duality of  $\mathfrak{sl}_n$ Toda field theory in order to obtain a strong/weak duality.
In subsection \ref{sec:Bchanging}, we examine the theory obtained in this procedure and observe $\mathfrak{sl}_{n|1}$-structure in the interaction terms.

\subsection{Gauge theory}

\label{BosonicGauging}

For our purpose, it is convenient to define the coset theory in the following way as in the case of $SL(2)/U(1)$ coset examined in \cite{Hikida:2008pe}.
Namely, we embed the theory which we wish to study in the combination of \(\mathfrak{sl}_n\) with a free boson $Y = Y_L + Y_R$ with \(Y(z) Y(w)= - \ln|z-w|^2 \), and fermionic ghosts \(b',c'\) arising from gauge fixing.  The action then takes the form 
\begin{equation}
    S=S^\text{sub}_k[\phi_l,\gamma,\beta]+\frac{1}{\pi}\int d^2w \left(\partial Y\overline{\partial}Y+\frac{1}{4}\sqrt{g}\mathcal{R} Q_{\tilde Y} {\tilde  Y} \right) +\frac{1}{\pi}\int d^2w (b'\overline{\partial}c'+\overline{b'}\partial \overline{c'}) \, .
\end{equation}
Here $S^\text{sub}_k[\phi_l,\gamma,\beta]$ is the action of the $\mathcal{W}^{(2)}_n$-theory given in \eqref{eq: action}. The background charge for $\tilde Y = Y_L - Y_R$, given by
\begin{align}
	Q_{\tilde Y} = - i \frac{n-2}{2}  \sqrt{\eta^2 +1 } \, , \label{QY}
\end{align}
arises since we are now dealing with a twisted sector as explained in appendix \ref{sec:twisting}.
The gauge fixing ghosts \(b',c'\) decouple from the rest of the theory, and make no notable contribution in the following work.  We consider the norm 0 diagonal Heisenberg field%
\footnote{In the BRST formulation adopted here, physical states are given by non-trivial elements of BRST cohomology. For the sector with BRST ghosts decoupled, the BRST closed condition becomes the condition of being regular with respect to $H$.}
\begin{equation}
    H(z)=J(z) + i\sqrt{\eta^2+1}\partial Y(z) \, . 
\end{equation}
We wish to consider the \(U(1)\) coset theory with respect to $H$.  To this end, we must find vertex operators in the above theory which are invariant under $H$.

For \(Y\), it is convenient to consider vertex operators which are acted on simply by $H$.  We then take the following form of vertex operators so that \( i\sqrt{\eta^2+1}\partial Y\) acts on \(V^Y_{m,\overline{m}}(w,\overline{w})\) as multiplication by \(\frac{m}{z-w}\)
\begin{align}
    V^Y_{m,\overline{m}}(z,\overline{z})=e^{i\frac{1}{\sqrt{\eta^2+1}}(mY_L-\overline{m}Y_R)} \, .
\end{align}
Vertex operators invariant under $H$ then take the  form 
\begin{equation}
    \Psi_{m,\overline{m}}^{\lambda_l}(z,\overline{z})=V^Y_{m,\overline{m}}(z,\overline{z})\Phi^{\lambda_l}_{m,\overline{m}}(z,\overline{z}) \, .
\end{equation}
In the case of $SL(2)/U(1)$ coset, it is known that the conservation of $U(1)$-charge is violated by the amount $k s /2$, where $ s = - N+2 , -N+1 , \ldots  , N-2$. This is related to the fact that the effects of spectral flow with winding number $s$ can be undone by $U(1)$-gauging,  but there is a shift of $U(1)$-eigenvalue by the amount $ks/2$. Performing the same procedure for the identity operator, we can obtain an identity operator in the coset model
consisting of the spectral flow operator 
\(v^s(\xi)\) of the $\mathfrak{sl}_2$ WZNW model \cite{Hikida:2008pe}, see also \cite{Creutzig:2020cmn} for more complicated examples.
Thus we also need to find the operator which we must pair with the spectral flow operator \(v^s(\xi)\) to make an identity operator in the coset theory. We recall that \(v^s\) acts on \(\beta\) as \(\beta_n\to \beta_{n+s}\) and on \(\gamma\) as \(\gamma_n \to \gamma_{n-s}\). Namely, we have
\begin{align}
:\beta\gamma:(z) v^s(\xi)&=\frac{1}{z-\xi}\sum_{n\in \mathbb{Z}}\left(\sum_{m<0}\beta_{m+s}\gamma_{n-s} z^{-m-n-1}+\sum_{m\geq0}\gamma_{n-s}\beta_{m+s}z^{-m-n-1}\right) \nonumber \\
&=\frac{1}{z-\xi}\sum_{n\in \mathbb{Z}}\left(\sum_{m<0}\beta_m\gamma_n z^{-m-n-1}+\sum_{m\geq0}\gamma_n\beta_mz^{-m-n-1}+\sum_{m=0}^{s-1}[\beta_m,\gamma_n]z^{-m-n-1}\right) \nonumber \\ &=\frac{1}{z-\xi}(:\beta\gamma:(z)+s) \, . \nonumber
\end{align} 
Recall also that \(v^s(\xi)\) acts by inserting the operator \(\exp ( \frac{s}{b}\phi^1(\xi) )\) at \(z = \xi\).  Then $J(z)$ acts on \(v^s(\xi)\) as  \begin{equation}\label{Jvs}
    J(z)v^s(\xi)=:\beta\gamma:(z)v^s(\xi)-\frac{1}{b}\partial \phi^1(z)v^s(\xi)
    =s(1+\eta^2)\frac{1}{z-\xi}v^s(\xi) \, . 
\end{equation}
From this, we can see that an identity operator in the coset theory takes the form
\begin{equation}\label{Bosonicgaugevs}
	V^Y_{-s ( \eta^2+1 ),-s ( \eta^2+1 ) }(\xi)v^s(\xi) \, . 
\end{equation} 
Noting the \(s,\xi\) subscript corresponding to insertion of the operator \(v^s(\xi)\), correlators in the gauge theory may be calculated as \begin{multline}\label{eq: Bosonic gauge correlator}
    \left \langle \prod_{\nu=1}^N\Psi^{\lambda_{\nu,l}}_{m_\nu,\overline{m}_\nu}(z_\nu)\right\rangle_{s,\xi}
    =\prod_{\nu=1}^N\left[N^{\lambda_{\nu,1}}_{m_\nu,\overline{m}_\nu}\int\frac{d^2\mu_\nu}{|\mu_\nu|^2}\mu_\nu^m\overline{\mu}_\nu^{\overline{m}_\nu}\right]\\
     \quad \times\left\langle V^Y_{-s (\eta^2+1),-s ( \eta^2+1)}(\xi)v^s(\xi)\prod_{\nu=1}^NV^Y_{m_\nu,\overline{m}_\nu}(z_\nu)V^\gamma_{\mu_\nu}(z_\nu)V_{\lambda_{\nu,l}}(z_\nu)\right\rangle \, . 
\end{multline}  
We require that
\begin{align}
	\sum_{\nu=1}^N m_\nu=(s +2 - n )(\eta^2+1)=\sum_{\nu=1}^N \overline{m}_\nu
\end{align}
to ensure that this correlator does not vanish due to charge neutrality, see \eqref{bosonicchargeneutrality} below.

\subsection{Shifting fields}
\label{sec:Bshifting}

We take the strategy of \cite{Hikida:2008pe,Creutzig:2020cmn} in order to obtain the dual theory.
For this we reduce the $\mathcal{W}_n^{(2)}$-theory into $\mathfrak{sl}_n$ Toda field theory as in \cite{Creutzig:2020ffn}.
We first integrate over \(\gamma,\overline{\gamma}\) which inserts delta functionals including \(\beta\) and \(\overline{\beta}\).  Following this with integration over \(\beta,\overline{\beta}\), we are left with  unpleasant factors in front of the \(S_1\) interaction term introduced in \eqref{screening}. The factors are
\begin{align} \label{betamu}
    \beta(w)=2\pi\sum_{\nu=1}^N\frac{\mu_\nu}{w-z_\nu} \, , \quad \overline{\beta}(w)=-2\pi\sum_{\nu=1}^N\frac{\overline{\mu}_\nu}{w-z_\nu} \, .
\end{align}
We note that due to the insertion of \(v^s(\xi)\), \(\beta\) and \(\overline{\beta}\) must have zeroes of order $s$ at \(\xi\), which impose the constraints
\begin{equation} \label{betaconst}
    0=\sum_{\nu=1}^N\frac{\mu_v}{(z_\nu-\xi)^n} 
\end{equation}
for \(n=0,...,s\).
We may remove these prefactors from the interaction term by shifting the zero-mode of \(\phi_1\).  To this end, it is convenient to write \(\beta(w), \overline{\beta}(w)\) in a product form.  We will denote the locations of remaining zeroes of \(\beta(w)\) by \(y_i\).  Then \(\beta\) can be written in the form 
\begin{align} \label{betay}
\beta(w)=u\frac{(w-\xi)^s\prod_{i=1}^{N-2-s}(w-y_i)}{\prod_{\nu=1}^N(w-z_\nu)}=u\mathcal{B}(w) 
\end{align}
with the multiple of \(u\).
The shift of \(\phi\) then takes the form
\begin{equation}
\begin{split}
    \varphi_1(w&) =\phi_1(w)+\frac{1}{b}\ln|u\mathcal{B}(w)|^2\\
    &=\phi_1(w)+\frac{1}{b}\left(s\ln|w-\xi|^2+\sum_{m=1}^{N-2-s}\ln|w-y_m|^2-\sum_{\nu=1}^N\ln|w-z_\nu|^2-\ln|u\rho(w)|^2\right) \, .
\label{Bosonic tildephi}
\end{split}
\end{equation}
The above field will be evaluated at \(w=z_\nu\), which leads to divergences. 
Here we adopt a regularization with the Weyl factor \(\rho(z)\) appearing in \eqref{metric} such that \(\lim_{z\to w}|w-z|^2=-\ln|\rho(z)|^2\).  Since the residue of \(u\mathcal{B}(w)\) around \(w=z_\nu\) is \(\mu_\nu\), we see that the  redefinition of the field \(\phi\) in \eqref{Bosonic tildephi} removes the \(|\mu|^{2 \lambda}\) prefactor from the \(\phi_1\) vertex operators in \eqref{vertexop1}, and similarly removes the \(\beta\overline{\beta}\) prefactor from the \(S_1\) screening operator. Namely, we have
\begin{align}
    V^1_{\lambda_\nu}(z_\nu)=|\mu_\nu|^{2 \lambda_{\nu,1}}e^{b\lambda_{\nu,1}\phi_1(z_\nu)}=e^{b\lambda_{\nu,1}\varphi_1(z_\nu)}=:V^{1'}_{\lambda_\nu}(z_\nu) \, , \quad 
    S_1=\beta\overline{\beta}e^{b\phi_1}=e^{b\varphi_1}=:S'_1 \, .
\end{align}
In the remainder of this work, we will use \(\sum_{l=1'}\) and \(\prod_{l=1'}\) to denote the replacement of a \(\phi_1\) quantity with the corresponding \(\varphi_1\) quantity.  We must now calculate the effect of this shift on the action.  
Noticing \eqref{curvature}, we have
\begin{multline}
\partial\overline{\partial}\varphi_1(w,\overline{w})=\partial\overline{\partial}\phi_1(w,\overline{w})+\frac{\pi}{b}\left(s\delta^2(w-\xi)+\sum_{i=1}^{N-2-s}\delta^2(w-y_i)-\sum_{\nu=1}^N \delta^2(w-z_\nu)\right)-\frac{1}{b}\partial\overline{\partial} \ln|\rho(w)|^2\\
=\partial\overline{\partial}\phi_1(w,\overline{w})+\frac{\pi}{b}\left(s\delta^2(w-\xi)+\sum_{i=1}^{N-2-s}\delta^2(w-y_i)-\sum_{\nu=1}^N \delta^2(w-z_\nu)\right)+\frac{2}{4b}\sqrt{g}\mathcal{R} \, .
\end{multline}
Using the fact that \(\partial \phi_1\overline{\partial}\phi_l\) and \( -\partial \overline{\partial}\phi_1\phi_l\) differ by a total derivative, we  further find
\begin{multline*}
-\frac{1}{\pi}\int d^2w \frac{1}{2}\sum_{l=2}^{n-1}\partial \phi_1\overline{\partial}\phi_lA^{-1}_{1l}=
    \frac{1}{\pi}\int d^2w\frac{1}{2}\sum_{l=2}^{n-1}\phi_l\partial \overline{\partial}\phi_1 A^{-1}_{1l}\\
    =\frac{1}{\pi}\int d^2w \frac{1}{2}\sum_{l=2}^{n-1}\phi_lA^{-1}_{1l}\left(\partial\overline{\partial}\varphi_1(w,\overline{w})-\frac{\pi}{b}\left(s\delta^2(w-\xi)+\sum_i\delta^2(w-y_i)-\sum_\nu\delta^2(w-z_\nu)\right)-\frac{1}{2b}\sqrt{g}\mathcal{R}\right)\\
    =-\frac{1}{\pi}\int d^2w \left[ \frac{1}{2}\sum_{l=2}^{n-1}A^{-1}_{1l}\left(\partial \varphi_1 \overline{\partial}\phi_l+\frac{1}{2b}\sqrt{g}\mathcal{R}\phi_l\right)-\frac{1}{2}\sum_{l=2}^{n-1}A^{-1}_{1l}\left(\frac{1}{b}s\phi_l(\xi)+\frac{1}{b}\sum_i\phi_l(y_i)-\frac{1}{b}\sum_\nu \phi_l(z_\nu)\right) \right] \, .
\end{multline*}
The \(\partial \phi_l\overline{\partial}\phi_1\) case is entirely similar, and makes the same additional contributions to the action, removing the \(\frac{1}{2}\) factor ahead of the additional insertions.  Here we introduce the new field
\begin{align}
    \varphi^1=\sum_{l=1'}^{n-1}A_{1l}^{-1}\phi_l \, , \quad V^{\varphi^1}_a=e^{\frac{1}{b}a\varphi^1} \, .
\end{align}
In summary, the shift of \(\phi_1 \to \varphi_1\) in \eqref{Bosonic tildephi} then does two things in the action: firstly, it forces the following field insertions in correlators
\begin{equation}
e^{-\frac{s}{b}\phi^1(\xi)}\prod_{i=1}^{N-2-s}e^{-\frac{1}{b}\varphi^1(y_i)}\prod_{\nu=1}^N\prod_{l=1'}^{n-1}e^{\frac{1}{b}A^{-1}_{1l}\phi_l(z_\nu)} \, .
\end{equation}
We note that \(\exp ( -\frac{s}{b}\phi^1(\xi) ) \) cancels the insertion from \(v^s(\xi)\), while \(\exp  ( \frac{1}{b}\phi^1(z_\nu))\) shifts the parameters that are already present in the correlators.  Secondly, the shift modifies the background charges such that \(\Delta Q_{\phi_l}=b^{-1}A_{1l}^{-1}\).  In particular, the existing background charge \(Q_{\phi^1}=b\) becomes \(Q_{\varphi^1}=b+b^{-1}\).  The \(\partial \phi_1\overline{\partial}\phi_1\) additionally contributes a numerical factor.

As in the previous analysis of \cite{Hikida:2008pe,Creutzig:2020cmn}, we also perform shift of the additional field as
\begin{align}
    \mathcal{Y}_L=Y_L-i\sqrt{\eta^2+1} \ln u \mathcal{B}(w) \, , 
\end{align}
where the shift \(\mathcal{Y}_R\) is given by the complex conjugate.  Then \(Y=Y_L+Y_R\) is shifted by \(-i\sqrt{\eta^2+1} \ln \mathcal{B}/\overline{\mathcal{B}}\) while \(\tilde{Y}=Y_L-Y_R\) is shifted by \(i\sqrt{\eta^2+1} \ln |u \mathcal{B}|^2\).  The results are as before, the \(\tilde{Y}\) background charge is shifted by $-i\sqrt{\eta^2+1}$, becoming 
\begin{align}
Q_{\tilde{\mathcal{Y}}}=-i  \frac{n}{2} \sqrt{\eta^2+1} \, , \label{bc}
\end{align}
and the following fields are inserted
\begin{align}
    e^{i\sqrt{\eta^2+1} s\tilde{Y}(\xi)}\prod_{i=1}^{N-2-s}e^{i\sqrt{\eta^2+1} \tilde{\mathcal{Y}}(y_i)}\prod_{\nu=1}^Ne^{-i\sqrt{\eta^2+1}\tilde{\mathcal{Y}}(z_\nu)} \, .
\end{align}
Again, the insertions at \(\xi\) cancel those required with the spectral flow operator \(v^s\).  The \(\mu,\overline{\mu}\) prefactors are also removed from the existing operators;
\begin{equation}
    (\mu_\nu)^m(\overline{\mu}_\nu)^{\overline{m}}V^Y_{m,\overline{m}}(z_\nu)=V^{\mathcal{Y}}_{m,\overline{m}}(z_\nu) \, .
\end{equation}

Through \eqref{betamu}, \eqref{betaconst} and \eqref{betay}, we changed the variables from $\mu_\nu$ to $y_i$ and its Jacobian is given by \cite{Ribault:2005wp,Ribault:2005ms,Hikida:2008pe}
\begin{align}
	\prod_{\mu =1}^N \frac{d^2 \mu_\nu}{|\mu_\nu|^2} \prod_{n=0}^s \delta^2 \left( \sum_{\nu} \frac{\mu_\nu}{(\xi - z_\nu)^n}\right) 
	= \frac{\prod_{\mu < \nu}^N |z_{\mu \nu}|^2 \prod_{i < j}^{N -2 - s} |y_{ij}|^2 }
	{\prod_{\nu=1}^N \prod_{i=1}^{N-2-s} |z_\nu - y_i|^2}  
	\frac{d^2 u}{|u|^{4 + 2 s}} \prod_{i=1}^{N-2-s} d^2 y_i \, . \label{Jacobian}
\end{align} 
We again obtained a numerical factor in front of the measure $d^2 u \prod_i d^2 y_i$, which actually cancels with those from the shifts of \(\phi\) and  \(Y\).  
The resulting correlators are 
\begin{multline}
    \left \langle \prod_{\nu=1}^N\Psi^{\lambda_{\nu,l}}_{m_\nu,\overline{m}_\nu}(z_\nu)\right\rangle_{s,\xi}=\prod_{i=1}^{N-2-s}\int\frac{d^2y_i}{(N-2-s)!}\prod_{\nu=1}^NN^{\lambda_{\nu,1}}_{m,\overline{m}}\\
    \times \left\langle \prod_{\nu=1}^NV_{\lambda_{\nu,l}+b^{-2}A_{1l}^{-1}}(z_\nu)V^{\mathcal{Y}}_{m_\nu-\eta^2-1,\overline{m}_\nu-\eta^2-1}(z_\nu)\prod_{i=1}^{N-2-s}V^{\varphi^1}_{-b^{-2}}(y_i)V^{\mathcal{Y}}_{\eta^2+1,\eta^2+1}(y_i)\right\rangle^{\tilde{S}} \, . \label{Nto2N-2-s}
\end{multline}
The division by $(N-2-s)!$ comes from the ambiguity in the map from $\mu_\nu$ to $y_i$, see, e.g, \cite{Hikida:2008pe}.
Here \(\tilde{S}\) is the earlier action with fields replaced with their corresponding shifted field, and with the new shifted background charges.

\subsection{Dual theory}
\label{sec:Bdual}

Now we have obtained a correspondence between a $N$-point function and a $(2N-2-s)$-point function as in \eqref{Nto2N-2-s}.
In order to show correspondence for correlation functions of two theories, we need to relate two $N$-point functions. This will be done in this subsection.

Before doing so, there is one important step to obtain a strong/weak duality.
Now a part of the right hand side of \eqref{Nto2N-2-s} is evaluated by $\mathfrak{sl}_n$ Toda field theory.
Here we apply the self duality of the field theory to all \(\phi_l\).  We may replace the screening charges \(S_l= \alpha e^{b\phi_l}\) for \(l=1',...,n-1\) with (see, e.g., \cite{Fateev:2007ab})
\begin{align}
\mathcal{S}_l=\tilde{\alpha} e^{\frac{1}{b}\phi_l} \, , \quad 
    \tilde{\alpha}=\gamma^{-1}(b^{-2})( \alpha\gamma(b^2))^{b^{-2}} \, ,
\end{align}
where \(\gamma(x)=\Gamma(x)/\Gamma(1-x)\).  The parameters of vertex operators need not be modified.  We will denote by \(\mathcal{S}\) the action \(\tilde{S}\) after this modification.

We must now rewrite the relation \eqref{Nto2N-2-s} as a correspondence of two $N$-point functions.
In order to ensure charge neutrality (so that the correlator is nondegenerate), we must impose the following condition on parameters of our vertex operators,
\begin{align}\label{bosonicchargeneutrality}
	\begin{aligned}
    0&=i\frac{1}{\sqrt{\eta^2+1}}\left(\sum_{\nu=1}^N(m_\nu-\eta^2-1)+\sum_{i=1}^{N-2-s}(\eta^2+1)\right)-2Q_{\tilde{\mathcal{Y}}} \\
    &=i\frac{1}{\sqrt{\eta^2+1}}\sum_{\nu=1}^Nm_\nu-i(2+s)\sqrt{\eta^2+1}-2Q_{\tilde{\mathcal{Y}}} \, .
    \end{aligned}
\end{align}
For this reason, the only term in \(\exp(V^{\mathcal{Y}}_{\eta^2}(y_i))\) making non-zero contribution will be that with power $(N-2-s)$.  We may then replace the extra inserted fields with this exponential and integrate over \(y_i\) as
\begin{align}
	\begin{aligned}
    & \left \langle \prod_{\nu=1}^N\Psi^{\lambda_{\nu,l}}_{m_\nu,\overline{m}_\nu}(z_\nu)\right\rangle_{s,\xi}
     =\int\frac{\prod_{i=1}^{N-2-s}d^2y_i}{(N-2-s)!}\prod_{\nu=1}^NN^{\lambda_{\nu,1}}_{m,\overline{m}}\\
    & \quad  \times\left\langle \prod_{\nu=1}^NV^{\mathcal{Y}}_{m_\nu-\eta^2-1,\overline{m}_\nu-\eta^2-1}(z_\nu)V_{\lambda_{\nu,l}+\frac{A^{-1}_{1l}}{b^2}}(z_\nu)\prod_{i=1}^{N-2-s}V^{\varphi^1}_{-b^{-2}}(y_i)V^{\mathcal{Y}}_{\eta^2+1,\eta^2+1}(y_i)\right\rangle^{\mathcal{S}} \\
     &=\pi^{N-2-s}\prod_{\nu=1}^NN^{\lambda_{\nu,1}}_{m,\overline{m}} \left\langle\prod_{\nu=1}^NV^{\mathcal{Y}}_{m_\nu-\eta^2-1,\overline{m}_\nu-\eta^2-1}(z_\nu)V_{\lambda_{\nu,l}+\frac{A^{-1}_{1l}}{b^2}}(z_\nu)\right.\\
     &\quad \left.
\times\exp\left(\frac{1}{\pi}\int d^2w\ V^{\mathcal{Y}}_{\eta^2+1,\eta^2+1}(w)V^{\varphi^1}_{-b^-2}(w)\right)\right\rangle^{\mathcal{S}} \, .
\end{aligned}
\end{align}
We arrive at the correlator of the form
\begin{multline}
     \left \langle \prod_{\nu=1}^N\Psi^{\lambda_{\nu,l}}_{m_\nu,\overline{m}_\nu}(z_\nu)\right\rangle_{s,\xi}\\
     =\pi^{N-2-s}\prod_{\nu=1}^NN^{\lambda_{\nu,1}}_{m,\overline{m}}
   \left\langle\prod_{\nu=1}^NV^{\mathcal{Y}}_{m_\nu-\eta^2-1,\overline{m}_\nu-\eta^2-1}(z_\nu)V_{\lambda_{\nu,l}+\frac{A^{-1}_{1l}}{b^2}}(z_\nu)\right\rangle^{\mathcal{S}-e^{-\frac{1}{b}\varphi^1+i\sqrt{\eta^2+1}\tilde{\mathcal{Y}}}} \, ,  \label{Bosonic degcalc}
\end{multline}
which corresponds to evaluation under the previous action with an additional interaction term \(\exp (-\frac{1}{b}\varphi^1+i\sqrt{\eta^2+1} \tilde{\mathcal{Y}} )\) \, . 

\subsection{Changing root systems in \(\mathfrak{sl}_{n|1}\) }
\label{sec:Bchanging}

In the previous subsection, we have obtained the correspondence of $N$-point functions of two theories. Moreover, as explained in appendix \ref{sec:Yalgebra}, the symmetry algebras of the two theories should match with each other. Therefore, we could say that we have already derived a strong/weak duality. 
However, in the case of FZZ-duality analyzed in \cite{Hikida:2008pe}, we need further steps to obtain a dual theory in the form of Sine-Liouville theory. In this subsection, we follow a similar procedure to bring the correlator correspondence into a nicer form.

We first defined the inner produce by  $p (z) q(w) \sim - (p , q) \ln (z - w)$ for some fields $p,q$.
We then introduce
\begin{align}\label{prerot bosonic interaction}
    q_0=-\frac{1}{b}\varphi^1+i\sqrt{\eta^2+1}\tilde{\mathcal{Y}} \, , \quad s_l=\frac{1}{b}\phi_l\quad (  l=1',...,n-1  ) \, ,
\end{align}
whose inner products are
\begin{align}
\label{eq: Bosonic slninner}
& (q_0 , q_0)=-1 \, , \quad  (q_0,s_l)=-\frac{1}{b^2}\delta_{1l} \, , \quad (s_l ,s_j)=\frac{1}{b^2}A_{lj} \, .
\end{align}
After our previous work, our interaction terms take the form 
\begin{align}\label{Bosonic preint}
    &Q_0=-e^{q_0} \, ,\quad \mathcal{S}_l=\tilde{\alpha}e^{s_l}\quad( l=1',...,n-1 ) \, .
\end{align}
As derived in \eqref{1stscreenings}, they corresponds to the screening operators for a free field realization of the desired algebra.
The inner products between the \(q_0\) and \(s_l\) take the form of a rescaling of the \(\mathfrak{sl}_{n|1}\) Cartan matrix by \(b^{-2}\), with one additional contribution in the top left corner 
\begin{equation}\label{eq: slngram}\left(\begin{array}{c|cccc}
     -1&-\frac{1}{b^2} &0 &0 &\hdots\\\hline
     -\frac{1}{b^2}&\frac{2}{b^2}& -\frac{1}{b^2}&0&\hdots\\
     0&-\frac{1}{b^2}&\frac{2}{b^2}&-\frac{1}{b^2}&\ddots\\
     0&0&-\frac{1}{b^2}&\frac{2}{b^2}&\ddots\\
     \vdots&\vdots&\ddots&\ddots&\ddots
\end{array}\right) \, .\end{equation}
This can be put into a slightly nicer form.  We note that our theory is invariant under the reflection
\begin{equation}\label{reflection}
    s_l\to s_l+ q_0 + \frac{2-2(s_l,q_0)}{(q_0,q_0)}q_0
\end{equation}
along with the modification of the constant ahead of the interaction term, 
see \cite{Hikida:2008pe} for details.
Note that only \(s_1\) is affected by this reflection, all other \(s_l\) are invariant as
\begin{align}
   s_1\to q_1=s_1-(1+2b^{-2})q_0=\frac{1}{b}\varphi_1-(1+2b^{-2})\left[-\frac{1}{b}\varphi^1+i\sqrt{\eta^2+1} \tilde{\mathcal{Y}}\right] \, .
\end{align}
More precisely speaking, we can change
\begin{align}
    \mathcal{S}_1\to Q_1=-\tilde{\alpha}\gamma \left(1+ 2 \eta ^2  \right)e^{q_1} \, ,
\end{align}
see appendix B of \cite{Hikida:2008pe} for the derivation of the factor.
In this form, the terms \(q_0,q_1\) and \(s_l\) have inner products 
\begin{equation}\label{eq: sl(n/1) gram}\left(\begin{array}{cc|cccc}
     -1&1+\frac{1}{b^2} &0 &0&0&\hdots\\[6pt]
    1+\frac{1}{b^2}&-1&-\frac{1}{b^2}&0&0&\hdots\\[6pt]\hline
     0&-\frac{1}{b^2}&\frac{2}{b^2}&-\frac{1}{b^2}&0&\hdots\\[6pt]
     0&0&-\frac{1}{b^2}&\frac{2}{b^2}&-\frac{1}{b^2}\\[6pt]
     0&0&0&-\frac{1}{b^2}&\ddots&\ddots\\[6pt]
     \vdots&\vdots&\vdots&&\ddots
\end{array}\right) \, .
\end{equation}
This gives a maximally odd description of the roots of \(\mathfrak{sl}_{n|1}\), again with an additional contribution in the top left.  We also apply the reflection to the vertex operators.  We define
\begin{align}
	\begin{aligned}
   & \lambda^L=\left(\frac{1}{b\eta^2}(\lambda_1+\eta^2),\ \hdots,b\left(\lambda_l-\frac{n-l}{n-1}\lambda_1\right),\hdots ,\ \frac{i}{\sqrt{\eta^2+1}}(m-\eta^2-1)\right)\label{Bosonic left coeff} \, , \\
&    \overrightarrow{\phi}^L=({\varphi^1}^L,\ \hdots,\ \phi_{l}^L,\hdots,\ \mathcal{Y}^L)
   \end{aligned}
\end{align}
and
\begin{align}
	\begin{aligned}
 &   \lambda^R=\left(\frac{1}{b\eta^2}(\lambda_1+\eta^2),\ \hdots,\ b\left(\lambda_l-\frac{n-l}{n-1}\lambda_1\right),\hdots,\ -\frac{i}{\eta}(\overline{m}-\eta^2-1)\right) \, , \\
  &  \overrightarrow{\phi}^R=({\varphi^1}^R,\ \hdots,\ \phi_{l}^R,\hdots,\ \mathcal{Y}^R) \, .
   \end{aligned}
\end{align}
So that vertex operators take the form 
\begin{equation}
    V_{\lambda}=\exp\left(\lambda^L\cdot\overrightarrow{\phi}^L+\lambda^R\cdot \overrightarrow{\phi}^R\right) \, .
\end{equation}
Notice that we are working with the basis \(\{\varphi^1,\hdots \phi_l,\hdots,\phi_{n-1}\} \) rather than \(\{\varphi_1,\hdots \phi_l,\hdots \phi_{n-1}\}\) as one might expect. The reflection then takes the form 
\begin{align}
    \lambda^L\to {\lambda^L}'=\lambda^L+ q_0 + \frac{2-2(\lambda^L,q_0)}{(q_0,q_0)}q_0^L=\lambda^L+\left(1-2[\lambda_1+m]\right)q_0^L \, . \label{Bosonic coeff reflection}
\end{align}
The right component is similar.  This also contributes a reflection coefficient ahead of the vertex operators, which removes the factor \(N^{\lambda_1}_{m,\overline{m}}\) up to a sign.  Resulting operators are 
\begin{equation}
    N^{\lambda_{1}}_{m,\overline{m}}V_{\lambda}\sim-e^{{\lambda^L}'\cdot \overrightarrow{\phi}^L+{\lambda^R}'\cdot \overrightarrow{\phi}^R} \, . 
\end{equation}
See \cite{Hikida:2008pe} again for the details.

We now redefine fields to simplify the forms of  \(\lambda_\nu\), \(q_0,q_1\) and \(s_l\).  
We define orthogonal fields \begin{align}
    \phi^1=(2\eta^2+1)\varphi^1-\frac{n-1}{n}2ib^{-1}\sqrt{\eta^2+1}\tilde{\mathcal{Y}} \, , \quad \tilde{Y}=-2ib^{-1}\sqrt{\eta^2+1}\varphi^1-(2\eta^2+1)\tilde{\mathcal{Y}} \label{rotation}\, .
\end{align}
Following these rotations, the background charges become \begin{align}
Q_{\phi^1}=b-(n-2)(b+b^{-1})\eta^2 \, , \quad Q_{\tilde{Y}}=-i\sqrt{\eta^2+1}\left[\frac{n-2}{2}-(n-2)(b^2+1)\eta^2\right] \, .
\label{rotbackgroundcharge}\end{align}
Using these fields, \(q_0\) and \(q_1\) take the forms 
\begin{align} \label{rint}
    q_0=\frac{1}{b}\phi^1-i\sqrt{\eta^2+1}\tilde{Y} \, , \quad 
    q_1=\frac{1}{b}\frac{1}{n-1}\phi^1+i\sqrt{\eta^2+1}\tilde{Y}-\frac{1}{b}\frac{n}{n-1}\sum_{l=2}^{n-1}A_{1l}^{-1}\phi_l \, .
\end{align}
We note that \(\phi^1\) and \(\tilde{Y}\) are mutually orthogonal, and orthogonal to all \(\phi_l,\ l=2,...,n-1\), and that \(\phi^1\) has norm \(\frac{n-1}{n}\).  With this in mind, we make one final field definition
 \begin{equation}
     \phi_1=\frac{n}{n-1}\left[\phi^1-\sum_{l=2}^{n-1}A_{1l}^{-1}\phi_l\right] \, .
 \end{equation}
Which is dual to $\phi^1$.  We note also that while \(\phi^1\) is orthogonal to all \(\phi_l\) for \(l=2,...,n-1\), this is no longer the case for \(\phi_1\).  Rather, we see that \(\phi_1(z)\phi_l(w)= \delta_{l2}\ln(z-w)\) for \( l=2,...,n-1\), and that \(\phi_1(z)\phi_1(w)= - 2\ln(z-w)\).  In particular, we see that OPEs of \(\phi_1,\phi_2,...,\phi_{n-1}\) are determined by the Cartan matrix of \(\mathfrak{sl}_{n-1}\), just as the OPEs of our original fields \(\phi_1,...,\phi_{n-1}\) were.  In these terms, two non-trivial interaction terms are given by  
\begin{align}
	\begin{aligned}
&	Q_0= e^{q_o} = e^{\frac{1}{b}\phi^1-i\sqrt{\eta^2+1}\tilde{Y}}  \, , \\
&	Q_1=-\tilde{\alpha}\gamma(1+2 \eta^2 )e^{q_1}=-c\tilde{\alpha}\gamma(1+2 \eta^2)e^{\frac{1}{b}\phi_1-\frac{1}{b}\phi^1+i\sqrt{\eta^2+1}\tilde{Y}} \, .
	\end{aligned}
\end{align}
All other interaction terms, \(S_l \) with \( l=2,...,n-1 \), are unaffected.  Suitable shift of the zero mode of \(\tilde{Y}\) leaves us with the following simple form for interaction terms
\begin{align}
	\begin{aligned}
    &Q_0=\kappa e^{\frac{1}{b}\phi^1-i\sqrt{\eta^2+1}\tilde{Y}} \, , \quad Q_1=\kappa e^{\frac{1}{b}\phi_1-\frac{1}{b}\phi^1+i\sqrt{\eta^2+1}\tilde{Y}} \, , \\  &\mathcal{S}_l=\tilde{\alpha}e^{\frac{1}{b}\phi_l}\quad (l=2,...,n-1) \, ,
   \end{aligned}
\end{align}
where we have defined
\begin{align}
    \kappa=\sqrt{\tilde{\alpha}\gamma(1+2 \eta^2)} \, .
\end{align}
There is one remaining subtlety before we arrive at the final form of our action.  We notice that following the rotation in \eqref{rotation}, the field $\tilde{Y}$ is not orthogonal to each of the $\phi^l=\sum_{l=1'}A_{1l}^{-1}\phi_l$ with $l=2,...,n-1$ as defined, as $\tilde{Y}$ is not orthogonal to $\varphi_1$.  We must then redefine the $\phi^l$ in terms of $\phi_1$, which is indeed orthogonal to $\tilde{Y}$.   We note that $\phi^l$ with $l=2,...,n-1$ appear only in the background charge, so we need only calculate the effect of this redefinition on the background charge.  To this end, we notice the following relationship following directly from the rotation  \eqref{rotation}
\begin{equation}
\varphi_1=\phi_1+2b^{-2}\phi^1-2ib^{-1}\sqrt{\eta^2+1}\tilde{Y} \, .
\end{equation}
Noting also that $\sum_{l=2}^{n-1}A_{1j}^{-1}=\sum_{l=2}^{n-1}\frac{n-l}{n}=b^2\eta^2\frac{n-2}{2}$, the effect on the background charge is then as follows;
\begin{align}
\sum_{l=2}Q_{\phi^l}\phi^l&=(b+b^{-1})\sum_{l=2}\sum_{j=1'}A_{lj}^{-1}\phi_l\\
&=(b+b^{-1})\eta^2(n-2)\phi^1- i (b^2+1)\eta^2(n-2)\sqrt{\eta^2+1}\tilde{Y}+(b+b^{-1})\sum_{l=2}\sum_{j=1}A_{lj}^{-1}\phi_l \, . \nonumber
\end{align}
Making a slight abuse of notation, we shall also denote the redefined dual fields by 
\begin{align}
\phi^l=\sum_{j=1}A_{lj}^{-1} \phi_j \, .
\end{align}
Then this redefinition simply results in a shift of the background charge in the $\phi^1$ and $\tilde{Y}$ directions.  We note that these shifts cancel with the additional terms in \eqref{rotbackgroundcharge}, so that we recover the background charges in \eqref{eq: action}.  Then our action takes the following form
\begin{multline}
S_k^P=\frac{1}{\pi}\int d^2w \left[  \frac{1}{2}\sum_{\substack{l,j=1}}^{n-1}\partial \phi_l\overline{\partial}\phi_jA_{lj}^{-1}+\frac{1}{4}\sqrt{g}\mathcal{R} \left(b \phi^1 + (b + b^{-1})\sum_{l=2}^{n-1} \phi^l \right) +Q_0+Q_1+\sum_{l=2}^{n-1} \mathcal{S}_l \right]\\
+\frac{1}{\pi}\int d^2w \left(\partial Y\overline{\partial}Y+\frac{1}{4}\sqrt{g}\mathcal{R} Q_{\tilde Y} \tilde Y \right)+\frac{1}{\pi}\int d^2w (b'\overline{\partial}c'+\overline{b'}\partial \overline{c'}) \, ,
\end{multline} 
where $Q_{\tilde Y}$ is given in \eqref{QY}.

Our final step is to write the vertex operators in terms of the new fields \(\phi_1\) and \(\tilde{Y}\).  Under this transformation, the vertex operator coefficients in \eqref{Bosonic left coeff} become the following
\begin{equation}
    \left(\frac{1}{b\eta^2}\lambda+b^{-1}(-1+2(\lambda_1+m)),\hdots, \frac{i}{\eta}(m+a)-i\sqrt{\eta^2+1}(-1+2(\lambda_1+m))\right) \, .
\end{equation}
Noting that the coefficients in \(q_0\) are ( \(b^{-1},\hdots, -i\sqrt{\eta^2+1}\) ), we see that additional coefficients from the reflection in  \eqref{Bosonic coeff reflection} cancel with the additional terms above, resulting in the coefficients 
\begin{equation}
    \lambda^L=\left(\frac{1}{b\eta^2}\lambda_1,\hdots,\ b\left(\lambda_l-\frac{n-l}{n-1}\lambda_1\right),\hdots,\ \frac{i}{\sqrt{\eta^2+1}}m\right) \, .
\end{equation}
The transformation is entirely similar for the right coefficients, with the only difference being the sign in the last two coefficients and replacement of \(m\) and \(a\) by \(\overline{m}\) and \(\overline{a}\), respectively.  We obtain the following
\begin{equation}
 \lambda^R=\left(\frac{1}{b\eta^2}\lambda_1,\hdots,\ b\left(\lambda_l-\frac{n-l}{n-1}\lambda_1\right),\hdots,\ -\frac{i}{\sqrt{\eta^2+1}}\overline{m} \right) \, .
 \end{equation}
After the final field redefinition from \(\phi'\) to \(\phi\), we have 
\begin{equation}V_\lambda=e^{\lambda^L\cdot \overrightarrow{\phi}^L+\lambda^R\cdot \overrightarrow{\phi}^R}\label{Bosonic final vertex op}\end{equation}
with
\begin{align}
&\lambda^L=\left(b\lambda_1,\hdots,\ b\lambda_l,\hdots,\ \frac{i}{\sqrt{\eta^2+1}}m\right) \, , \quad 
 \overrightarrow{\phi}^L=(\phi^L,\ \phi^L_2,\hdots,\ \phi^L_{n-1},\ Y^L) \, ,  \\
&     \lambda^R=\left(b\lambda_1,\hdots,\ b\lambda_l,\hdots,\ -\frac{i}{\sqrt{\eta^2+1}}\overline{m}\right) \, , \quad 
     \overrightarrow{\phi}^R=(\phi^R,\phi_2^R,\hdots,\ \phi_{n-1}^R,Y^R) \, . \label{Bosonic final right coeff}
 \end{align} 
 We note that these are exactly the coefficients that we began with in the \(\mathfrak{sl}_{n}\) subregular \(\mathcal{W}\)-algebra with additional boson \(Y\).  We come to the final equality
 \begin{equation}
      \left \langle \prod_{\nu=1}^N\Psi^{\lambda_{\nu,l}}_{m_\nu,\overline{m}_\nu}(z_\nu)\right\rangle_{s,\xi}=\mathcal{N}\left\langle \prod_{\nu=1}^N V_{\lambda_\nu}(z_\nu)\right\rangle^{S^P_k} \, , 
 \end{equation}
where \(V_{\lambda_\nu}\) are determined as in \eqref{Bosonic final vertex op}-\eqref{Bosonic final right coeff} and \(\mathcal{N}=(-1)^{N-s}\pi^{N-2-s}\kappa^s\).

\section{Fermionic duality}
\label{sec:Fduality}

In this section, we derive another series of strong/weak duality by adding additional pair of free fermions and gauging a diagonal Heisenberg algebra.  The analysis is quite analogous to the previous case.  Even so, we believe that it is worth demonstrating as the symmetry algebra in this case is the regular $\mathcal{W}$-superalgebra of $\mathfrak{sl}_{n|1}$.

The strategy is the same as in the previous case.
In the next subsection, we realize the coset theory by embedding the products of several theories.
In subsection \ref{sec:Fshifting}, we perform the shift of fields but now including the (bosonized version) of fermionic fields.
In subsection \ref{sec:Fdual}, we obtain the dual theory by applying the self-duality of Toda field theory and interpreting extra inserted fields as an interaction term.
In subsection \ref{sec:Fchanging}, we rewrite the dual theory in a nice form as for $\mathcal{N}=2$ Liouville field theory in this case.

\subsection{Gauge theory}
\label{Fermionic Gauging}

The theory which we wish to study may be embedded in the combination of \(\mathfrak{sl}_n\) with a free boson $Y$ with \(Y(z) Y(w)=-\ln|z-w|^2\), fermionic ghosts \(b,c\) for introducing additional fields%
\footnote{In the case of fermionic FZZ-duality with $n=2$, the theories possess the symmetry of $\mathcal{N}=2$ superconformal algebra. However, for other cases with $n > 2$, there is no supersymmetry in the usual sense.}
and fermionic ghosts \(b',c'\) arising from gauge fixing.  It is convenient to bosonize the ghosts \(b,c\) as 
\begin{align}
b=e^{iZ_L} \, , \quad c=e^{-iZ_L} 
\end{align}
with \( Z_L(z)Z_L(w)=-\ln(z-w) \).   
The action then takes the form 
\begin{multline}
    S=S^\text{sub}_k[\phi_l,\gamma,\beta]+\frac{1}{\pi}\int d^2w \left(\partial Y\overline{\partial}Y+\frac{1}{4}\sqrt{g}\mathcal{R} Q_{\tilde Y} \tilde Y \right)
+\frac{1}{\pi}\int d^2 w\partial Z\overline{\partial}Z +\frac{1}{\pi}\int d^2w (b'\overline{\partial}c'+\overline{b'}\partial \overline{c'})
\end{multline}
with $S^\text{sub}_k[\phi_l,\gamma,\beta]$ defined in \eqref{eq: action}. The background charge for $\tilde Y$ is given by
\begin{align}
	Q_{\tilde Y} = - i \frac{n-2}{2} ( \eta + \eta^{-1} ) \, , \label{sQY}
\end{align}
see appendix \ref{sec:twisting}.
The gauge fixing ghosts \(b',c'\) decouple from the rest of the theory, and make no notable contribution in the following work.  We consider the norm 0 diagonal Heisenberg field
\begin{equation}
    H(z)=J(z) - i\partial Z(z) + i\eta\partial Y(z) \, . \label{sHei}
\end{equation}
We wish to consider the \(U(1)\) coset theory with respect to $H$.  To this end, we must find vertex operators in the above theory which are invariant under $H$.

For \(Y\) and \(Z\), it is convenient to consider vertex operators which are acted on simply by $H$.  We then take the following form of vertex operators so that \(- i\partial Z(z)+i\eta\partial Y\) acts on \(V^Y_{m,\overline{m}}(w,\overline{w})V^Z_{m,\overline{m}}(w,\overline{w})\) as multiplication by \( m / ( z-w )\) as
\begin{align}
    V^Y_{m+a,\overline{m}+\overline{a}}(z,\overline{z})=e^{i\frac{1}{\eta}((m+a)Y_L-(\overline{m}+\overline{a})Y_R)} \, , \quad V^Z_{a,\overline{a}}=e^{i(aZ_L-\overline{a}Z_R)} \, .
\end{align}
Vertex operators invariant under $H$ then take the form 
\begin{equation}
    \Psi_{m,\overline{m}}^{\lambda_l,a}(z,\overline{z})=V^Y_{m+a,\overline{m}+\overline{a}}(z,\overline{z})V^Z_{a,\overline{a}}(z,\overline{z})\Phi^{\lambda_l}_{m,\overline{m}}(z,\overline{z}) \, .
\end{equation}
All that remains is finding the operators which we must pair with the spectral flow operator \(v^s(\xi)\) in order to make an identity operator in the coset theory.
Noticing that \(J(z)\) acts on \(v^s(\xi)\) just as in the bosonic case, see \eqref{Jvs},
an identity operator can be  constructed as 
\begin{equation}\label{Fermionic gaugevs}V^Y_{-s\eta^2,-s\eta^2}(\xi)V^Z_{s,s}(\xi)v^s(\xi) \, . \end{equation}
Noting the \(s,\xi\) subscript corresponding to insertion of the operator \(v^s(\xi)\), correlators in the supersymmetric gauge theory may be calculated as \begin{multline}\label{eq: Fermionic gauge correlator}
    \left \langle \prod_{\nu=1}^N\Psi^{\lambda_{\nu,l},a_\nu}_{m_\nu,\overline{m}_\nu}(z_\nu)\right\rangle_{s,\xi}
    =\prod_{\nu=1}^N\left[N^{\lambda_{\nu,1}}_{m_\nu,\overline{m}_\nu}\int\frac{d^2\mu_\nu}{|\mu_\nu|^2}\mu_\nu^m\overline{\mu}_\nu^{\overline{m}_\nu}\right]\\
    \times\left\langle V^Y_{-s\eta^2,-s\eta^2}(\xi)V^Z_{s,s}(\xi)v^s(\xi)\prod_{\nu=1}^NV^Y_{m_\nu+a_\nu,\overline{m}_\nu+\overline{a}_\nu}(z_\nu)V^Z_{a_\nu,\overline{a}_\nu}(z_\nu)V^\gamma_{\mu_\nu}(z_\nu)V_{\lambda_{\nu,l}}(z_\nu)\right\rangle \, .
\end{multline}  We require that \(\sum_{\nu}(m_\nu+a_\nu)=\eta^2  s - (n- 2)(1 + \eta^2)=\sum_{\nu}(\overline{m}_\nu+\overline{a}_\nu)\) to ensure that this correlator does not vanish due to charge neutrality, see \eqref{Fermionic chargeneutrality} below.

\subsection{Shifting fields}
\label{sec:Fshifting}

The earliest steps are entirely similar to the bosonic case.  We integrate first over \(\gamma\) and \(\overline{\gamma}\), and then over \(\beta\) and \(\overline{\beta}\), replacing the \(\beta\overline{\beta}\) prefactor in \(S_1\) with \(|u\mathcal{B}(w)|^2\).  We again perform the field redefinition
\begin{equation}
    \varphi_1(w)=\phi_1(w)+\frac{1}{b}\ln |u\mathcal{B}(w)|^2 \, . 
\end{equation}
This removes the prefactors both from \(S_1\) and from \(V^1_{\lambda_\nu}\)
\begin{align}
    V^1_{\lambda_\nu}(z_\nu)=|\mu_\nu|^{2 \lambda_{\nu,1}}e^{b\lambda_{\nu,1}\phi_1(z_\nu)}=e^{b\lambda_{\nu,1}\varphi_1(z_\nu)}=:V^{1'}_{\lambda_\nu}(z_\nu) \, , \quad  
    S_1=\beta\overline{\beta}e^{b\phi_1}=e^{b\varphi_1}=:S'_1 \, .
\end{align}
As before, the shift \(\phi_1 \to \varphi_1 \) forces the following insertions in correlators
\begin{equation}
e^{-\frac{s}{b}\phi^1(\xi)}\prod_{i=1}^{N-2-s}e^{-\frac{1}{b}\varphi^1(y_i)}\prod_{\nu=1}^N\prod_{l=1'}^{n-1}e^{\frac{1}{b}A^{-1}_{1l}\phi_l(z_\nu)} \, .
\end{equation}
These cancel the insertion from \(v^s(\xi)\), shift the parameters that are already present in the correlators, and insert new operators at \(y_i\).  The background charge is modified such that \(Q_{\phi^1}=b\) becomes \(Q_{\varphi^1}=b+b^{-1}\).  
A numerical factor also appears.

In this case we also perform the shifts both in \(Y\) and \(Z\) as
\begin{align}
    \mathcal{Y}_L=Y_L-i\eta \ln u \mathcal{B}(w) \, , \quad \mathcal{Z}_L=Z_L+i\ln u \mathcal{B}(w) \, , 
\end{align}
where \(\mathcal{Y}_R\) and \(\mathcal{Z}_R\) are given by the complex conjugate.  Then \(Y=Y_L+Y_R\) is shifted by \(-i\eta \ln \mathcal{B}/\overline{\mathcal{B}}\) while \(\tilde{Y}=Y_L-Y_R\) is shifted by \(i\eta\ln |u \mathcal{B}|^2\), and similarly for \(Z=Z_L+Z_R\) and \(\tilde{Z}=Z_L-Z_R\).  The results are as before, \(\tilde{Y}\) and \(\tilde{Z}\) obtain background charges of \(Q_{\tilde{\mathcal{Y}}}=-i\frac{n-2}{2} (\eta + \eta^{-1}) - i \eta \) and  \ \(Q_{\tilde{\mathcal{Z}}}=i \), respectively, and lead to insertions of the fields 
\begin{align}
    e^{i\eta s\tilde{Y}(\xi)}\prod_{i=1}^{N-2-s}e^{i\eta \tilde{\mathcal{Y}}(y_i)}\prod_{\nu=1}^Ne^{-i\eta\tilde{\mathcal{Y}}(z_\nu)} \, , \quad e^{-i s \tilde{Z}(\xi)}\prod_{i=1}^{N-2-s}e^{-\tilde{\mathcal{Z}}(y_i)}\prod_{\nu=1}^Ne^{\tilde{\mathcal{Z}}(z_\nu)} \, .
\end{align}
Again, the insertions at \(\xi\) cancel those required with the spectral flow operator \(v^s\).  The \(\mu,\overline{\mu}\) prefactors are also removed from the existing operators as
\begin{equation}
    (\mu_\nu)^m(\overline{\mu}_\nu)^{\overline{m}}V^Y_{m+a,\overline{m}+\overline{a}}(z_\nu)V^Z_{a,\overline{a}}(z_\nu)=V^{\mathcal{Y}}_{m+a,\overline{m}+\overline{a}}(z_\nu)V^{\mathcal{Z}}_{a,\overline{a}}(z_\nu) \, .
\end{equation}
The resulting correlators are 
\begin{multline}
    \left \langle \prod_{\nu=1}^N\Psi^{\lambda_{\nu,l},a_\nu}_{m_\nu,\overline{m}_\nu}(z_\nu)\right\rangle_{s,\xi}=\prod_{i=1}^{N-2-s}\int\frac{d^2y_i}{(N-2-s)!}\prod_{\nu=1}^NN^{\lambda_{\nu,1}}_{m,\overline{m}}\\
    \times \left\langle \prod_{\nu=1}^NV_{\lambda_{\nu,l}+b^{-2}A_{1l}^{-1}}(z_\nu)V^{\mathcal{Y}}_{m_\nu+a_\nu-\eta^2,\overline{m}_\nu+\overline{a}_\nu-\eta^2}(z_\nu)V^{\mathcal{Z}}_{a_\nu+1,\overline{a}_\nu+1}(z_\nu)\prod_{i=1}^{N-2-s}V^{\varphi^1}_{-b^{-2}}(y_i)V^{\mathcal{Y}}_{\eta^2,\eta^2}(y_i)V^{\mathcal{Z}}_{-1,-1}(y_i)\right\rangle^{\tilde{S}} \, ,  \label{FNtoN-2-s}
\end{multline}
where \(\tilde{S}\) is the earlier action with fields replaced with their corresponding shifted field, and with the new shifted background charges. The integration over $y_i$ comes from the change of variables from $\mu_\nu$ and $y_i$, and the Jacobian due to the changes is given by \eqref{Jacobian}. In particular, the numerical factors arising from the shifts of $\phi_1 , Y , Z$ total cancel out that of the Jacobian.

\subsection{Dual theory}
\label{sec:Fdual}

We would like to write \eqref{FNtoN-2-s} as a correspondence between $N$-point functions.
Before that, we first apply the self duality of this field theory to all \(\phi_l\) in order to obtain a strong/weak duality.
We may replace the screening charges \(S_l= \alpha e^{b\phi_l}\) for \(l=1',...,n-1\) with 
\begin{align}
\mathcal{S}_l=\tilde{\alpha} e^{\frac{1}{b}\phi_l} \, , \quad 
    \tilde{\alpha}=\gamma^{-1}(b^{-2})( \alpha\gamma(b^2))^{b^{-2}} \, , 
\end{align}
where \(\gamma(x)=\Gamma(x)/\Gamma(1-x)\).  The parameters of vertex operators need not be modified.  We will denote by \(\mathcal{S}\) the action \(\tilde{S}\) after this modification.

We then examine the correlation functions in the right hand side of \eqref{FNtoN-2-s}.
In order to ensure charge neutrality (so that the correlator is nondegenerate), we must impose the following condition on parameters of our vertex operators
\begin{equation}\label{Fermionic chargeneutrality}
    0=i\frac{1}{\eta}\left(\sum_{\nu=1}^N(m_\nu+a_\nu-\eta^2)+\sum_{i=1}^{N-2-s}\eta^2\right)-2Q_{\tilde{\mathcal{Y}}}=i\frac{1}{\eta}\sum_{\nu=1}^N(m_\nu+a_\nu)-i(2+s)\eta-2Q_{\tilde{\mathcal{Y}}} \, .
\end{equation}
For this reason, the only term in \(\exp(V^{\mathcal{Y}}_{\eta^2}(y_i))\) making non-zero contribution will be that with power $(N-2-s)$.  We may then replace the inserted fields with this exponential and integrate over \(y_i\) as
\begin{multline*}
     \left \langle \prod_{\nu=1}^N\Psi^{\lambda_{\nu,l},a_\nu}_{m_\nu,\overline{m}_\nu}(z_\nu)\right\rangle_{s,\xi}
     =\int\frac{\prod_{i=1}^{N-2-s}d^2y_i}{(N-2-s)!}\prod_{\nu=1}^NN^{\lambda_{\nu,1}}_{m,\overline{m}}\\
     \times\left\langle \prod_{\nu=1}^NV^{\mathcal{Y}}_{m_\nu+a_\nu-\eta^2,\overline{m}_\nu+\overline{a}_\nu-\eta^2}(z_\nu)V^{\mathcal{Z}}_{a_\nu+1,\overline{a}_\nu+1}(z_\nu)V_{\lambda_{\nu,l}+\frac{A^{-1}_{1l}}{b^2}}(z_\nu)\prod_{i=1}^{N-2-s}V^{\varphi^1}_{-b^{-2}}(y_i)V^{\mathcal{Y}}_{\eta^2,\eta^2}(y_i)V^{\mathcal{Z}}_{-1,-1}(y_i)\right\rangle^{\mathcal{S}} \\
     =\pi^{N-2-s}\prod_{\nu=1}^NN^{\lambda_{\nu,1}}_{m,\overline{m}} \left\langle\prod_{\nu=1}^NV^{\mathcal{Y}}_{m_\nu+a_\nu-\eta^2,\overline{m}_\nu+\overline{a}_\nu-\eta^2}(z_\nu)V^{\mathcal{Z}}_{a_\nu+1,\overline{a}_\nu+1}(z_\nu)V_{\lambda_{\nu,l}+\frac{A^{-1}_{1l}}{b^2}}(z_\nu)\right.\\\left.
     \times\exp\left(\frac{1}{\pi}\int d^2w\ V^{\mathcal{Y}}_{\eta^2,\eta^2}(w)V^{\mathcal{Z}}_{-1,-1}(w)V^{\varphi^1}_{-b^-2}(w)\right)\right\rangle^{\mathcal{S}} \, .
\end{multline*}
We thus arrive at the correlator of the  form 
\begin{multline}
     \left \langle \prod_{\nu=1}^N\Psi^{\lambda_{\nu,l},a_\nu}_{m_\nu,\overline{m}_\nu}(z_\nu)\right\rangle_{s,\xi}\\
     =\pi^{N-2-s}\prod_{\nu=1}^NN^{\lambda_{\nu,1}}_{m,\overline{m}}
   \left\langle\prod_{\nu=1}^NV^{\mathcal{Y}}_{m_\nu+a_\nu-\eta^2,\overline{m}_\nu+\overline{a}_\nu-\eta^2}(z_\nu)V^{\mathcal{Z}}_{a_\nu+1,\overline{a}_\nu+1}(z_\nu)V_{\lambda_{\nu,l}+\frac{A^{-1}_{1l}}{b^2}}(z_\nu)\right\rangle^{\mathcal{S}-e^{-\frac{1}{b}\varphi^1+i\eta\tilde{\mathcal{Y}}-i\tilde{\mathcal{Z}}}}\label{Fermionic degcalc} \, ,
\end{multline}
which corresponds to evaluation under the previous action with an additional interaction term \(- \exp ( -\frac{1}{b}\varphi^1+i\eta \tilde{\mathcal{Y}}-i\tilde{\mathcal{Z}} ) \).

\subsection{Changing root systems in \(\mathfrak{sl}_{n|1}\) }
\label{sec:Fchanging}

As in the case of bosonic dualities, we can stop here, since we have already obtained a strong/weak duality.
However, in the $n=2$ case, it is convenient to take further steps to bring the dual theory into the form of $\mathcal{N}=2$ Liouville field theory \cite{Creutzig:2010bt}. Therefore, we will follow a similar procedure here.

We first introduce 
\begin{align}
    q_0=-\frac{1}{b}\varphi^1+i\eta\tilde{\mathcal{Y}}-i\tilde{\mathcal{Z}} \, , \quad s_l=\frac{1}{b}\phi_l\quad  (  l=1',...,n-1 ) \, .
\end{align}
After our previous work, our interaction terms take the form 
\begin{align}\label{Fermionic preint}
    Q_0=-e^{q_0} \, , \quad \mathcal{S}_l=\tilde{\alpha}e^{s_l}\quad ( l=1',...,n-1 ) \, .
\end{align}
Inner products between the \(q_0\) and \(s_l\) are the same as the bosonic case, taking the form \eqref{eq: slngram} of a rescaling of the \(\mathfrak{sl}_{n|1}\) Cartan matrix by \(b^{-2}\), with one additional contribution in the top left corner.  This time however, we see that this additional contribution in the top left corner is entirely from the bosonized fermion \(\mathcal{Z}\).  We again perform the reflection \eqref{reflection}, recovering the reflected Gram matrix \eqref{eq: sl(n/1) gram}
\begin{align}
    s_1\to q_1=s_1-(1+2b^{-2})q_0 \, , \quad \mathcal{S}_1\to Q_1=-\tilde{\alpha}\gamma(1+ 2 \eta^2 )e^{q_1} \, .
\end{align}
This gives a maximally odd description of the roots of \(\mathfrak{sl}_{n|1}\), again with an additional contribution in the top left from the fermions.  

We then consider the vertex operators inserted.
We define
\begin{align}
	\begin{aligned}
  &  \lambda^L=\left(\frac{1}{b\eta^2}(\lambda_1+\eta^2),\ \hdots,b\left(\lambda_l-\frac{n-l}{n-1}\lambda_1\right),\hdots ,\ \frac{i}{\eta}(m+a-\eta^2),\ i(a+1)\right) \, , \label{Fermionic left coeff}
  \\
 &   \overrightarrow{\phi}^L=({\varphi^1}^L,\ \hdots,\ \phi_{l}^L,\hdots,\ \mathcal{Y}^L,\mathcal{Z}^L) \, , 
\end{aligned} 
\end{align}
and
\begin{align}
\begin{aligned}
 &   \lambda^R=\left(\frac{1}{b\eta^2}(\lambda_1+\eta^2),\ \hdots,\ b\left(\lambda_l-\frac{n-l}{n-1}\lambda_1\right),\hdots,\ -\frac{i}{\eta}(\overline{m}+\overline{a}-\eta^2),\ -i(\overline{a}+1)\right) \, , \\
&    \overrightarrow{\phi}^R=({\varphi^1}^R,\ \hdots,\ \phi_{l}^R,\hdots,\ \mathcal{Y}^R,\mathcal{Z}^R)
\end{aligned}
\end{align}
so that vertex operators take the form 
\begin{equation}
    V_{\lambda}=\exp\left(\lambda^L\cdot\overrightarrow{\phi}^L+\lambda^R\cdot \overrightarrow{\phi}^R\right) \, .
\end{equation}
We will proceed with reflection on the vertex operators as in the bosonic case
\begin{align}
    \lambda^L\to {\lambda^L}'=\lambda^L+\frac{1-2(\lambda^L,q_0)}{(q_0,q_0)}q_0^L=\lambda^L+\left(1-2[\lambda_1+m]\right)q_0^L \, . \label{Fermionic coeff reflection}
\end{align}
The right component is similar.  This also contributes a reflection coefficient ahead of the vertex operators, which removes the factor \(N^{\lambda_1}_{m,\overline{m}}\) up to a sign.  Resulting operators are 
\begin{equation}
    N^{\lambda_{1}}_{m,\overline{m}}V_{\lambda}\sim-e^{{\lambda^L}'\cdot \overrightarrow{\phi}^L+{\lambda^R}'\cdot \overrightarrow{\phi}^R} \, .
\end{equation}
We now redefine fields to simplify the forms of  \(\lambda_\nu\), \(q_0,q_1\) and \(s_l\).  We first define  
\begin{align}
    \tilde{\mathcal{Y}}^+=\frac{1}{\sqrt{\eta^2+1}}\left(\tilde{\mathcal{Y}}+\eta\tilde{\mathcal{Z}}\right) \, , \quad \tilde{\mathcal{Y}}^-=\frac{1}{\sqrt{\eta^2+1}}\left(\eta\tilde{\mathcal{Y}}-\tilde{\mathcal{Z}}\right) \, .
\end{align}
We see that the background charges of these fields are
\begin{align}
	Q_{\tilde{\mathcal{Y}}^+}= - i \frac{(n-2) \sqrt{1 + \eta^2}}{ 2 \eta} \, , \quad 
	Q_{\tilde{\mathcal{Y}}^-}=-i \frac{n}{2}  \sqrt{\eta^2+1} \, .
\end{align}
Note that we may identify $Q_{\tilde{\mathcal{Y}}^-}$ in the fermionic case with $Q_{\tilde{\mathcal{Y}}}$ in \eqref{bc} in the bosonic case. With this rotation, \(q_0\) takes the form 
\begin{equation}
    q_0=-\frac{1}{b}\varphi^1+i\sqrt{\eta^2+1} \tilde{\mathcal{Y}}^- \, .
\end{equation}
In doing this redefinition, we have recovered the form of the bosonic interaction terms in \eqref{prerot bosonic interaction}.  We may then proceed as in the bosonic case, by defining orthogonal fields \begin{align}
    \phi^1=(2\eta^2+1)\varphi^1-\frac{n-1}{n}2ib^{-1}\sqrt{\eta^2+1}\tilde{\mathcal{Y}}^- \, , \quad \tilde{Y}^-=-2ib^{-1}\sqrt{\eta^2+1}\varphi^1-(2\eta^2+1)\tilde{\mathcal{Y}}^{-} \, .
\end{align}
Using these fields, \(q_0\) and \(q_1\) take the forms 
\begin{align}
    q_0=\frac{1}{b}\phi^1-i\sqrt{\eta^2+1}\tilde{Y}^- \, , \quad
    q_1=\frac{1}{b}\frac{1}{n-1}\phi^1+i\sqrt{\eta^2+1}\tilde{Y}^--\frac{n}{n-1}\sum_{l=2}^{n-1}A_{1l}^{-1}\phi_l\, .
\end{align}
Background charges for these fields are
\begin{align}
Q_{\phi^1}=b-(n-2)(b+b^{-1})\eta^2 \, , \quad Q_{\tilde{Y}^-}=-i\sqrt{\eta^2+1}\left[\frac{n-2}{2}-(n-2)(b^2+1)\eta^2\right] \, . \label{fermionicrotbackgroundcharge}\end{align} 
As in the bosonic case, the redefinitions 
\begin{equation}
\phi^l=\sum_{j=1}^{n-1}A_{lj}^{-1}\phi_j\end{equation}
shift the background charges in the $\phi^1$ and $\tilde{Y}^-$ directions, leaving us with 
\begin{align}
	Q_{\phi^1}=b \, , \quad Q_{\tilde{Y}^-}=-i\sqrt{\eta^2+1}\left(\frac{n-2}{2}\right) \, .
\end{align}
We again define  \begin{equation}
     \phi_1=\frac{n}{n-1}\left[\phi^1-\sum_{l=2}^{n-1}A_{1l}^{-1}\phi_l\right] \, .
 \end{equation}
which is dual to $\phi^1$ and, along with the $\phi_l$ for $l=2,\ldots , n-1$, transforms in the $\mathfrak{sl}_{n-1}$ Cartan matrix, and apply the inverse of our earlier field definition to \(\tilde{\mathcal{Y}}^+\) and our new field \(\tilde{Y}^-\)
\begin{align}
    \tilde{Y}=\frac{1}{\sqrt{\eta^2+1}}\left(\tilde{\mathcal{Y}}^++\eta\tilde{Y}^-\right) \, , \quad \tilde{Z}=\frac{1}{\sqrt{\eta^2+1}}\left(\eta \tilde{\mathcal{Y}}^+-\tilde{Y}^-\right) \, .
\end{align}
The resulting form of interaction terms is as follows 
\begin{align}
    q_0=\frac{1}{b}\phi^1-i\eta \tilde{Y}+i\tilde{Z} \, , \quad q_1=\frac{1}{b}\phi_1-\frac{1}{b}\phi^1+i\eta\tilde{Y}-i\tilde{Z}
\end{align}
and background charges become
\begin{align}
Q_{\tilde{Y}}= - i \frac{n-2}{2} ( \eta + \eta^{-1} )  \, , \quad Q_{\tilde{Z}}= 0 \, . \end{align}
Note that \(\phi^1,\tilde{Y}\) and \(\tilde{Z}\) are all mutually orthogonal, and orthogonal to all \(\phi_l\) with \( l=2,...,n-1\), and that \(\phi^1\) and \(\tilde{Z}\) have norms \(\frac{n-1}{n}\) and 1, respectively.  Then \(i\tilde{Z}\) and \(-i\tilde{Z}\) again play the role of bosonized \(bc\) fermions, with 
\begin{align}
    b=e^{i\tilde{Z}} \, , \quad c=e^{-i\tilde{Z}} \, .
\end{align}
Applying this result, along with suitable shift of the zero mode of \(\tilde{Y}\), we are left with interaction terms
\begin{align}
    Q_0=b\kappa e^{\frac{1}{b}\phi^1-i\eta\tilde{Y}} \, , \quad Q_1=c\kappa e^{\frac{1}{b}\phi_1-\frac{1}{b}\phi^1+i\eta\tilde{Y}}\, , \quad \kappa=\sqrt{\tilde{\alpha}\gamma(1+ 2 \eta^2)} \, .
\end{align}
All other interaction terms are unaffected,
\begin{align}
   \mathcal{S}_l=\tilde{\alpha}e^{b\phi_l}\quad (  l=2,...,n-1 ) \, . 
\end{align}
Notice, in this form, the exponential part of the interaction terms transform as roots of \(\mathfrak{sl}_{n|1}\) with no additional contributions.  As in the bosonic case, we see that OPEs of \(\phi_1,\phi_2,...,\phi_{n-1}\) are determined by the Cartan matrix of \(\mathfrak{sl}_{n}\).  Then our action takes the form 
\begin{multline}
S_k^{P}=\frac{1}{\pi}\int d^2w \left[  \frac{1}{2}\sum_{\substack{l,j=1}}^{n-1}\partial \phi_l\overline{\partial}\phi_jA_{lj}^{-1}+\frac{1}{4}\sqrt{g}\mathcal{R} \left(b \phi^1 + (b + b^{-1} )\sum_{l=2}^{n-1} \phi^l \right)+Q_0+Q_1+\sum_{l=2}^{n-1} \mathcal{S}_l \right] \\
+\frac{1}{\pi}\int d^2w \left(\partial Y\overline{\partial}Y+\frac{1}{4}\sqrt{g}\mathcal{R} Q_{\tilde Y} \tilde Y \right)+\frac{1}{\pi}\int d^2 w(b\overline{\partial}c+\overline{b}\partial c)
+\frac{1}{\pi}\int d^2w (b'\overline{\partial}c'+\overline{b'}\partial \overline{c'}) \, .
\end{multline} 

Our final step is to write the vertex operators in terms of the new fields \(\tilde{Y}\) and \(\tilde{Z}\).  We note that the three field redefinitions performed previously taking fields \(\varphi^1,\tilde{\mathcal{Y}}, \tilde{\mathcal{Z}}\) to fields \(\phi^1,\tilde{Y}, \tilde{Z}\) take the form of the single field redefinition 
\begin{equation}
    \left(\begin{array}{ccc}
         2\eta^2+1&-2ib\eta^3&2ib\eta^2\\
         -2i\eta b^{-1}&-2\eta^2+1&2\eta\\
         2ib^{-1}&2\eta&-1 
    \end{array}\right) \, .
\end{equation}
Note also that this transformation is self-inverse.  Then under this transformation, the vertex operator coefficients in \eqref{Fermionic left coeff} become the following in the basis \({\phi'}^L,\tilde{Y}^L\) and \(\tilde{Z}^L\)
\begin{equation}
    \left(\begin{array}{c}
         \frac{1}{b\eta^2}\lambda+b^{-1}(-1+2(\lambda_1+m))\\\vdots \\\frac{i}{\eta}(m+a)-i\eta(-1+2(\lambda_1+m))\\ia+i(-1+2(\lambda_1+m))
    \end{array}\right) \, .
\end{equation}
Noting that the coefficients in \(q_0\) are (\(b^{-1},\hdots, -i\eta, i\)), we see that additional coefficients from the reflection in  \eqref{Fermionic coeff reflection} cancel with the additional terms above, resulting in the coefficients 
\begin{equation}
    \lambda^L=\left(\frac{1}{b\eta^2}\lambda_1,\hdots,\ b\left(\lambda_l-\frac{n-l}{n-1}\lambda_1\right),\hdots,\ \frac{i}{\eta}(m+a), ia\right) \, .
\end{equation}
The transformation is entirely similar for the right coefficients, with the only difference being the sign in the last two coefficients and replacement of \(m\) and \(a\) by \(\overline{m}\) and \(\overline{a}\), respectively.  We obtain the following
\begin{equation}
 \lambda^R=\left(\frac{1}{b\eta^2}\lambda_1,\hdots,\ b\left(\lambda_l-\frac{n-l}{n-1}\lambda_1\right),\hdots,\ -\frac{i}{\eta}(\overline{m}+\overline{a}), -i\overline{a}\right) \, .
 \end{equation}
After the final field redefinition from \(\phi'\) to \(\phi\), we have 
\begin{equation}V_\lambda=e^{\lambda^L\cdot \overrightarrow{\phi}^L+\lambda^R\cdot \overrightarrow{\phi}^R}\label{Fermionic final vertex op}\end{equation}
with
\begin{align}
&\lambda^L=\left(b\lambda_1,\hdots,\ b\lambda_l,\hdots,\ \frac{i}{\eta}(m+a),ia\right) \, , \quad 
 \overrightarrow{\phi}^L=(\phi^L,\ \phi^L_2,\hdots,\ \phi^L_{n-1},\ Y^L,Z^L) \, , \\
 &    \lambda^R=\left(b\lambda_1,\hdots,\ b\lambda_l,\hdots,\ -\frac{i}{\eta}(\overline{m}+\overline{a}),-i\overline{a}\right) \, , \quad 
     \overrightarrow{\phi}^R=(\phi^R,\phi_2^R,\hdots,\ \phi_{n-1}^R,Y^R, Z^R) \, .\label{Fermionic final right coeff}
 \end{align} 
 Again, these are exactly the coefficients that we began with in the \(\mathfrak{sl}_{n}\) subregular \(\mathcal{W}\)-algebra, with additional boson \(Y\) and bosonized fermion \(Z\).  We come to the final equality,
 \begin{equation}
      \left \langle \prod_{\nu=1}^N\Psi^{\lambda_{\nu,l},a_\nu}_{m_\nu,\overline{m}_\nu}(z_\nu)\right\rangle_{s,\xi}=\mathcal{N}\left\langle \prod_{\nu=1}^N V_{\lambda_\nu}(z_\nu)\right\rangle^{S^P_k} \, ,
 \end{equation}
  where \(V_{\lambda_\nu}\) are determined as in \eqref{Fermionic final vertex op}-\eqref{Fermionic final right coeff} and \(\mathcal{N}=(-1)^{N-s}\pi^{N-2-s}\kappa^s\).

\subsection*{Acknowledgements}

The work of TC is supported by NSERC Grant Number RES0048511.
The work of YH is supported by JSPS KAKENHI Grant Number 19H01896, 21H04469.

\appendix

 \section{Twisting the energy momentum tensor}
 \label{sec:twisting}

 The subregular $\mathcal{W}$-algebra of $\mathfrak{sl}_n$ can be described by $n-1$ free bosons $\phi_l$ and one $(\beta,\gamma)$-system with screening charges \eqref{screening} as explained in the main context, see \cite{Feigin:2004wb,Creutzig:2020vbt}.
 However, the original algebra is realized with $(\beta,\gamma)$ with conformal weight not equal to $(1,0)$. See,  e.g., section 2.1 and section 4.1 of \cite{Creutzig:2020ffn}.
 It is convenient to move to a sector where $(\beta,\gamma)$ has conformal weight $(1,0)$ by twisting the energy momentum tensor. After the twisting, we can apply our approach developed in \cite{Hikida:2007tq,Creutzig:2020ffn}. For the case of Bershadsky-Polyakov algebra, this is equivalent to move to the Ramond sector \cite{Arakawa:2010sc}.
 
 Denoting the energy momentum tensor and $U(1)$ current by $T,J$, the twist is obtained by
 \begin{align}
 	T ' = T - \frac{n-2}{2} \partial J \, .
 \end{align}
This essentially reproduces the expressions for $n=2,3,4$ in \cite{Creutzig:2020ffn}.
 The original energy momentum tensor is given by
 \begin{align}
 	T = - \frac12 \partial \phi^i A_{ij} \partial \phi^j +  (b+ b^{-1}) \sum_{l=1}^n \partial^2 \phi^l - \frac{n}{2 b} \partial^2 \phi^1 - ( \partial \beta  ) \gamma + \frac{n}{2} \partial (\beta \gamma) \, ,
 \end{align}
 whose central charge is
 \begin{align}
 	c = 	\frac{(k (n-1)-(n-2) n) \left(k (n-2) n-(n-3) n^2-1\right)}{k-n} \, .
 \end{align}
 This reproduces eq.~(1.1) of \cite{Feigin:2004wb} after replacing $k$ by $-k$.
 The OPE between $T '$ and $J$ is now changed as
 \begin{align}
 	T ' (z) J(w) \sim  - 2 \frac{n-2}{2} \frac{1 + \eta^2}{(z -w)^3} + \frac{J(w)}{(z-w)^2} + \frac{\partial J(w)}{z-w} \, . \label{3rd}
 \end{align}
 The coefficient of the third order pole corresponds to the anomalous charge conservation, see, e.g., appendix C of \cite{Eberhardt:2019ywk}. This, in particular, implies that $Y$ has background charge  as in \eqref{QY}
 for the bosonic case and as in \eqref{sQY} for the fermionic case.

 \section{Free field realizations of  $Y_{1,0,n}$-algebra}
 \label{sec:Yalgebra}

 According to \cite{Gaiotto:2017euk}, the $Y_{0,1,n}[\psi]$-algebra can be realized by a $U(1)$ coset of subregular W-algebra as
 \begin{align}
 	Y_{0,1,n}[\psi] = \frac{W_{n-1,1}[U(n)_\psi]}{U(1)_\psi} = \frac{W_{n-1,1}[SU(n)_{\psi - n}] \times U(1)_{n \psi}}{U(1)_\psi}  \, .
 \end{align}
 We may change the parameter $\psi$ as $\psi = - k + n$. 
 Via its construction of brane junction, we can see that the algebra should be dual to $Y_{1,0,n}[\psi^{-1}]$-algebra. In the following, we find free field realizations of the algebra as in subsection 2.2 of \cite{Creutzig:2020cmn} by applying the method of \cite{Prochazka:2018tlo}, see also \cite{BFM,Litvinov:2016mgi} for previous works.
 
 For this, we introduce $1+n$ free bosons $\phi^{(1)} , \phi^{(3)}_j$ with $ j = 1,2,\ldots,n$. Non-trivial OPEs are given by 
 \begin{align}
 	\begin{aligned}
& 	\phi^{(1)} (z) \phi^{(1)} (w)\sim - \frac{1}{h_2 h_3} \ln (z - w)  \, , \\
 &	\phi^{(3)}_j (z) 	\phi^{(3)}_{j '} (w) \sim - \frac{1}{h_1 h_2}  \delta_{j,j'} \ln (z - w)  = \delta_{j,j'} \ln ( z - w)  \, , 
    \end{aligned}
 \end{align}
 where
 \begin{align}
 	h_1 = i \sqrt{k - n} \, , \quad  h_2 = \frac{i}{\sqrt{k -n}} \, , \quad h_3 = - i \frac{k-n+1}{\sqrt{k -n}}  \, .
 \end{align}
 There are many free field realizations and they are organized by the order of the free bosons.
 Below we examine two examples among them.
 
 First we study the free field realization corresponding to the order $\phi^{(1)} \phi^{(3)}_1 \cdots \phi^{(3)}_n$.
 The screening operators are given by
 \begin{align}
 	V_1 = e^{- h_1 \phi^{(3)}_1 + h_3 \phi^{(1)}} \, , \quad V_l = e^{- h_1  \phi^{(3)}_l + h_1 \phi^{(3)}_{l-1} }
 \end{align}
 with $l=2,3,\ldots ,n$. As in \cite{Creutzig:2020cmn}, we would like to decouple a Heisenberg algebra generated by
 $\phi^{(1)} + \sum_{l=1}^n \phi^{(3)}_l$. Free fields orthogonal to the direction are
 \begin{align}
 	\phi_j = i (\phi^{(3)}_j - \phi^{(3)}_{j+1}) \, , \quad 
 	Y = \frac{- i}{\sqrt{1 + \eta^2}} \left( h_3 \phi^{(1)} - \frac{h_1}{n} \sum_{j=1}^n \phi^{(3)}_j \right)  \, . \label{phiY}
 \end{align}
 Here we have defined
 \begin{align}
 	1 + \eta^2 = - \frac{h_3}{h_2} + \frac{h_1^2}{n} = \frac{n k - k - n^2 + 2 n}{n} \, .
 \end{align}
 The OPEs are
 \begin{align}
 	\phi_i (z) \phi_j (w) \sim - A_{ij} \ln (z - w) \, , \quad Y(z) Y(w) \sim - \ln (z - w) \, .
 \end{align}
 In terms of $\phi_j,Y$, the screening operators are rewritten as
 \begin{align}
 	V_1 = e^{ - \frac{1}{b} \phi^1 + i \sqrt{1 + \eta^2}Y} \, , \quad V_l = e^{\frac{1}{b} \phi_{l-1}} \, . \label{1stscreenings}
 \end{align}
 The screening operators reproduce \eqref{Bosonic preint} and \eqref{prerot bosonic interaction}.
 
 Next we examine the free field realization corresponding to the order $\phi^{(3)}_1  \phi^{(1)} \phi^{(3)}_2 \cdots \phi^{(3)}_n$.
 The screening operators are written as
 \begin{align}
 	V_1 = e^{- h_3 \phi^{(1)} + h_1 \phi^{(3)}_1 } \, , \quad	V_2 = e^{- h_1 \phi^{(3)}_2 + h_3 \phi^{(1)}} \, , \quad  V_l = e^{- h_1  \phi^{(3)}_l + h_1 \phi^{(3)}_{l-1} }
 \end{align}
 with $l=3,4,\ldots ,n$. 
 In terms of $\phi_j,Y$ defined in \eqref{phiY}, the screening operators become
 \begin{align}
 	V_1 = e^{ \frac{1}{b} \phi^1 - i \sqrt{1 + \eta^2}Y} \, , \quad V_2 = e^{ \frac{1}{b} \frac{1}{n-1} \phi^1 + i \sqrt{1 + \eta^2}Y - \frac{n}{n-1} \sum_{j=2}^{n-1} \frac{n-l}{n} \phi_j} \, , \quad V_l = e^{\frac{1}{b} \phi_{l-1}} \label{2ndscreenings}
 \end{align}
 with $l=3,4,\ldots , n$.
 The screening operators reproduce those with \eqref{rint}.

%\bibliographystyle{JHEP}
%\bibliography{FS}

\providecommand{\href}[2]{#2}\begingroup\raggedright\endgroup

\end{document}